\documentclass[aps,prc,showpacs,preprintnumbers,superscriptaddress,showkeys,twocolumn]{revtex4}

\usepackage{latexsym}
\usepackage{amssymb}
\usepackage{amsmath}
\usepackage{graphicx}
\usepackage{bm}
\usepackage{epsf}
\usepackage{epstopdf}

\newcommand{\MeV}{\mathrm{MeV}}
\newcommand{\fm}{\mathrm{fm}}
\usepackage[usenames]{color}
\usepackage[normalem]{ulem}  %

\newcommand{\comment}[1]{}

\renewcommand\sout{\bgroup \color{red} \ULdepth=-.5ex \ULset}

\begin{document}
\preprint{YITP-18-56}

\title{Effects of quark-matter symmetry energy on hadron-quark coexistence in neutron-star matter}
\author{Xuhao Wu} ~\email{wuhaobird@gmail.com}
\affiliation{School of Physics, Nankai University, Tianjin 300071, China}
\affiliation{Yukawa Institute for Theoretical Physics, Kyoto University, Kyoto 606-8502, Japan}
\author{Akira Ohnishi} ~\email{ohnishi@yukawa.kyoto-u.ac.jp}
\affiliation{Yukawa Institute for Theoretical Physics, Kyoto University, Kyoto 606-8502, Japan}
\author{Hong Shen}~\email{shennankai@gmail.com}
\affiliation{School of Physics, Nankai University, Tianjin 300071, China}

\begin{abstract}
We examine the effects of the isovector-vector coupling and hypercharge-vector coupling in quark matter
 on hadron-quark coexistence in neutron-star matter.
The relativistic mean field theory with the TM1 parameter set
and an extended TM1 parameter set are used to describe hadronic matter,
and the Nambu-Jona-Lasinio model with scalar, isoscalar-vector, isovector-vector and hypercharge-vector
couplings is used to describe deconfined quark matter.
The hadron-quark phase transition is constructed via the Gibbs conditions for phase equilibrium.
The isovector-vector and hypercharge-vector couplings in quark matter enhance the symmetry energy
and hypercharge symmetry energy in neutron-star matter,
while their effects are found to be suppressed at high densities by the strange quarks.
As a result, the hadron-quark mixed phase shrinks with only isovector-vector coupling and moves
to higher density with isovector-vector and hypercharge-vector couplings.
The maximum mass of neutron-star increases slightly with isovector-vector and hypercharge-vector couplings.
\end{abstract}

\maketitle


\section{Introduction}
\label{sec:1}
Existence of the first-order phase transition in dense neutron-star matter is of primary
interest in nuclear physics and compact star astrophysics.
In the inner core region of massive neutron stars,
the baryon density may reach $(5-10)n_0$
($n_0\simeq 0.15~\fm^{-3}$), and
the chiral and deconfinement hadron-quark phase transition may occur~\cite{Glen01,Heis00,Webe05}.
At present, the sign problem prevents us from predicting the properties of cold dense matter
in the first principles calculations such as the lattice QCD Monte Carlo simulations,
then phenomenological approaches as well as experimental and observational data are necessary
to explore the inner part of neutron stars.

One of the theoretical approaches to study the phase transition is
to apply chiral effective models such as the Nambu-Jona-Lasinio (NJL) model.
The first-order phase transition can occur at the baryon chemical potential
$\mu_b=(1000-1200)~\MeV$ using the chiral effective models.
Unfortunately, the order of the transition depends on the parameters,
and the equation of state (EOS) of nuclear matter at low densities
is not well described by the NJL model.
Thus it is more realistic to consider the coexistence
of hadronic and quark matter
in order to predict the transition density.
The QCD phase transition signal during the core collapse supernovae
was investigated by using a relativistic mean field (RMF) model for hadronic matter
and a bag model for quark matter~\cite{Nakazato08,Sagert:2008ka},
and the early collapse~\cite{Nakazato08}
or the second shock~\cite{Sagert:2008ka}
was found to signal the transition to quark matter.
In this case, the transition is necessarily of the first order
and the transition density strongly depends on the bag constant.
The hadron-quark coexistence is investigated also by using
RMF for hadronic matter
and the NJL model for quark matter~\cite{Benic:2014jia,Wu17,Pereira16}.
In RMF, the isovector-vector meson ($\rho$) plays an important role
to control the symmetry energy in hadronic matter,
and the coupling has been carefully chosen to explain the properties
of finite nuclei.
By comparison, the isovector-vector coupling in quark matter has been considered
less carefully.
The isovector-vector coupling constant ($G_3$) was chosen to be 
$G_3=1.5G_0$ in Refs.\ \cite{Wu17,Rehb96} and $G_3=G_0$ in Ref.\ ~\cite{Ueda13},
where $G_0$ is the isoscalar-vector coupling,
while the isovector-vector coupling was ignored in Ref.\ \cite{Benic:2014jia}.
This difference comes from the two independent types
of the chiral $\mathrm{SU}(N_f)$ vector coupling terms,
$(\bar{q}\gamma_\mu q)^2$
and 
$\sum_\alpha \left[(\bar{q}\gamma_\mu \lambda_\alpha q)^2
+(\bar{q}i\gamma_\mu\gamma_5 \lambda_\alpha q)^2\right]$ \cite{Chu16}.
The second type includes the isovector-vector coupling terms
and gives rise to symmetry energy in quark matter,
the energy increase from the $u$ and $d$ quark imbalance.
Since the symmetry energy in nuclear matter is known to affect
the neutron-star properties such as radii,
it is expected to be important also in the hadron-quark coexistence.

In this work, we examine the role of the isovector-vector coupling
in quark matter on the hadron-quark coexistence in neutron-star matter.
For this purpose, we apply RMF for hadronic matter
and the three-flavor NJL model for quark matter,
and we compare the coexistence density region with and without
the isovector-vector coupling term in NJL.
A finite isovector-vector coupling in quark matter enhances
the symmetry energy in quark matter,
which characterizes the increase of the energy per baryon
from unbalanced $u$ and $d$ quark densities.
In addition to this {\em isospin} symmetry energy,
the hypercharge symmetry energy appears with $N_f=3$
and controls the $s$ quark contribution.
We also examine the effects of nuclear matter symmetry energy slope $L$.
We use the TM1 parameter set ($L=110.8$ MeV)~\cite{Suga94}
and an extended TM1 parameter set (TM1e)~\cite{Bao14b},
where the symmetry energy at the density of $n_b=0.11$ fm$^{-3}$ is fixed
and the symmetry energy slope is tuned to be $L=50~\MeV$.
We find that the quark-matter symmetry energy
increases the starting density of the hadron-quark coexistence
and enhances the maximum mass of neutron stars.
The quark-matter symmetry energy effects on the hadron-quark coexistence
are suppressed by the $s$ quarks
and are smaller than those of the nuclear matter symmetry energy.

There are two comments in order.
First, it should be noted that Pereira {\em et al}. have already
discussed the effects of the isovector-vector coupling of quarks
on the hadron-quark coexistence, and have found that
the isovector-vector coupling pushes up the starting density of the coexistence
for a given value of the isoscalar-vector coupling~\cite{Pereira16}.
One of the differences of the present work and Ref.~\cite{Pereira16} is
in the hadronic matter EOS.
For hadronic matter,
they adopt the NL3$\omega\rho$ parameter set
which derives stiffer EOS at high density than TM1 and TM1e,
then most of the parameter sets predict
the $1.4 M_\odot$ neutron-star radii larger than the range,
$10\,\mathrm{km} \lesssim R_{1.4} \lesssim 13.6\,\mathrm{km}$~\cite{Annala2017},
constrained by the gravitational wave observation 
from a binary neutron-star merger event~\cite{LIGO2017}.
Thus it would be valuable to examine TM1e,
which predicts smaller neutron-star radii as shown later
in Table \ref{tab:2coex}.
Second, we do not consider here hyperon admixture in neutron stars.
If hyperons $\Lambda, \Sigma$, and $\Xi$ are taken into account
in a standard way in RMF, the neutron-star maximum mass is known
to become much smaller, and many hyperonic matter EOSs cannot
support $2 M_\odot$, as shown for example in Ref.~\cite{Ishizuka2008}.
One of the ways to avoid this {\em hyperon puzzle}
is to consider additional repulsion for hyperons at high densities,
then the hyperon fractions would be smaller and their effect may not be large.
Therefore the role of hyperons should be limited,
while we need to introduce additional couplings in hadronic matter.

This article is organized as follows.
In Sec.~\ref{sec:models}, we introduce
the RMF model and the NJL model with the isovector-vector coupling for hadronic and quark matter,
and we briefly describe the Gibbs conditions used as the equilibrium conditions
in the hadron-quark mixed phase.
In Sec. \ref{sec:results}, we show
the numerical results of the hadron-quark coexistence in neutron-star matter
and discuss the impact of the isovector-vector coupling.
Section \ref{sec:summary} is devoted to a summary.

\section{Hadronic matter, quark matter and hadron-quark coexistence}
\label{sec:models}

\subsection{Hadronic matter}
\label{sec:RMF}
We adopt the RMF theory to describe the hadronic phase, in which baryons
interact by exchanging the isoscalar scalar ($\sigma$),
isoscalar vector ($\omega$), and isovector vector
($\rho$) mesons.
These mesons are treated as classical field
under the mean field approximation.
For neutron-star matter, we use the Lagrangian given as
\begin{eqnarray}
\label{eq:LRMF}
\mathcal{L}_{\rm{RMF}} & = & \sum_{i=p,n}\bar{\psi}_i
\bigg \{i\gamma_{\mu}\partial^{\mu}-\left(M+g_{\sigma}\sigma\right)
\notag\\
&&-\gamma_{\mu} \left[g_{\omega}\omega^{\mu} +\frac{g_{\rho}}{2}\tau_a\rho^{a\mu}
\right]\bigg \}\psi_i  \notag \\
&& +\frac{1}{2}\partial_{\mu}\sigma\partial^{\mu}\sigma -\frac{1}{2}%
m^2_{\sigma}\sigma^2-\frac{1}{3}g_{2}\sigma^{3} -\frac{1}{4}g_{3}\sigma^{4}
\notag \\
&& -\frac{1}{4}W_{\mu\nu}W^{\mu\nu} +\frac{1}{2}m^2_{\omega}\omega_{\mu}%
\omega^{\mu} +\frac{1}{4}c_{3}\left(\omega_{\mu}\omega^{\mu}\right)^2  \notag
\\
&& -\frac{1}{4}R^a_{\mu\nu}R^{a\mu\nu} +\frac{1}{2}m^2_{\rho}\rho^a_{\mu}%
\rho^{a\mu}
\notag\\&&
+\Lambda_{\rm{v}} \left(g_{\omega}^2
\omega_{\mu}\omega^{\mu}\right)
\left(g_{\rho}^2\rho^a_{\mu}\rho^{a\mu}\right) \notag\\
&& +\sum_{l=e,\mu}\bar{\psi}_{l}
  \left( i\gamma_{\mu }\partial^{\mu }-m_{l}\right)\psi_l,
\end{eqnarray}
which contains the contributions of baryons  ($n$ and $p$) and leptons  ($e$ and $\mu$).
$W^{\mu\nu}$ and $R^{a\mu\nu}$ are the antisymmetric field
tensors for $\omega^{\mu}$ and $\rho^{a\mu}$, respectively.
The parameters in the Lagrangian are usually determined by fitting nuclear matter
saturation properties and ground-state properties of finite nuclei. We use the TM1 parameter
set~\cite{Suga94} and an extended TM1 parameter set~\cite{Bao14b}, referred to as
the TM1e parameter set in later discussions. In TM1e, the symmetry energy
slope parameter is tuned to be $L=50~\MeV$
at saturation density, as listed in Table~\ref{tab:para}.
For the homogeneous matter system, the meson field equations have the following form:
\begin{eqnarray}
&&m_{\sigma }^{2}\sigma +g_{2}\sigma ^{2}+g_{3}\sigma
^{3}=-g_{\sigma }\left( n_{p}^{s}+n_{n}^{s}\right) ,
\label{eq:eqms} \\
&&m_{\omega }^{2}\omega +c_{3}\omega^{3}
+2\Lambda_{\rm{v}}g^2_{\omega}g^2_{\rho}{\rho}^2 \omega
=g_{\omega}\left( n_{p}+n_{n}\right) ,
\label{eq:eqmw} \\
&&m_{\rho }^{2}{\rho}
+2\Lambda_{\rm{v}}g^2_{\omega}g^2_{\rho}{\omega}^2{\rho}
=\frac{g_{\rho }}{2}\left(n_{p}-n_{n}\right) ,
\label{eq:eqmr}
\end{eqnarray}%
where $n_i^s$ and $n_i$ represent the scalar and vector densities
of the $i$th baryon ($i=n, p$), respectively.
The equations of motion for nucleons give the standard relations between the densities
and chemical potentials,
\begin{eqnarray}
\mu_{p} &=& {\sqrt{\left( k_{F}^{p}\right)^{2}+{M^{\ast }}^{2}}}+g_{\omega}\omega +\frac{g_{\rho }}{2}\rho,
\label{eq:mup} \\
\mu_{n} &=& {\sqrt{\left( k_{F}^{n}\right)^{2}+{M^{\ast }}^{2}}}+g_{\omega}\omega -\frac{g_{\rho }}{2}\rho,
\label{eq:mun}
\end{eqnarray}%
where $M^{\ast}=M+g_{\sigma}\sigma$ is the effective nucleon mass,
and $k_{F}^{i}$ is the Fermi momentum of species $i$, which is related to the number density 
by $n_i=\left(k_{F}^{i}\right)^3/3\pi^2$. For neutron-star matter in $\beta$ equilibrium,
the chemical potentials satisfy the relations $\mu_{p}=\mu_{n}-\mu_{e}$ and $\mu_{\mu}=\mu_{e}$,
where the chemical potentials of leptons are given by $\mu_{l}=\sqrt{{k_{F}^{l}}^{2}+m_{l}^{2}}$.
In neutron-star matter, the total energy density and pressure are given by
\begin{eqnarray}
\varepsilon &=&\sum_{i=p,n}\frac{1}{\pi^2}
     \int_{0}^{k^{i}_{F}}{\sqrt{k^2+{M^{\ast}}^2}}k^2dk   \nonumber \\
&& + \frac{1}{2}m^2_{\sigma}{\sigma}^2+\frac{1}{3}{g_2}{\sigma}^3
     +\frac{1}{4}{g_3}{\sigma}^4
 + \frac{1}{2}m^2_{\omega}{\omega}^2+
     \frac{3}{4}{c_3}{\omega}^4  \nonumber  \\
&&   + \frac{1}{2}m^2_{\rho}{\rho}^2
     + 3{\Lambda}_{\textrm{v}}\left(g^2_{\omega}{\omega}^2\right)
     \left(g^2_{\rho}{\rho}^2\right)
     + \varepsilon_l,
     \label{eq:ehp}
\end{eqnarray}
\begin{eqnarray}
P &=& \sum_{i=p,n}\frac{1}{3\pi^2}\int_{0}^{k^{i}_{F}}
      \frac{1}{\sqrt{k^2+{M^{\ast}}^2}}k^4dk    \nonumber  \\
&& - \frac{1}{2}m^2_{\sigma}{\sigma}^2-\frac{1}{3}{g_2}{\sigma}^3
     -\frac{1}{4}{g_3}{\sigma}^4     \nonumber \\
&& + \frac{1}{2}m^2_{\omega}{\omega}^2
      +\frac{1}{4}{c_3}{\omega}^4
      + \frac{1}{2}m^2_{\rho}{\rho}^2
\nonumber\\
&&    + \Lambda_{\textrm{v}}\left(g^2_{\omega}{\omega}^2\right)
      \left(g^2_{\rho}{\rho}^2\right)
      + P_l,
           \label{eq:php}
\end{eqnarray}
where $\varepsilon_l$ and $P_l$ ($l=e, \mu$) are the energy density and the pressure
from leptons, respectively.

\begin{table*}[tbp]
\caption{Parameters in the TM1 and TM1e parameter sets.
The masses are given in the unit of MeV.}
\begin{center}
\begin{tabular}{lcccccccccccc}
\hline\hline
Model   &$L$(MeV) &$M$  &$m_{\sigma}$  &$m_\omega$  &$m_\rho$  &$g_\sigma$  &$g_\omega$
        &$g_\rho$ &$g_{2}$ (fm$^{-1}$) &$g_{3}$ &$c_{3}$  &$\Lambda_{\rm{v}}$\\
\hline
TM1     &110.8 &938.0  &511.198  &783.0  &770.0  &10.0289  &12.6139  &9.2644
        &$-$7.2325   &0.6183   &71.3075   &0\\
\hline
TM1e    &50   &938.0  &511.198  &783.0  &770.0  &10.0289  &12.6139  &12.2413
        &$-$7.2325   &0.6183   &71.3075   &0.0327\\

\hline\hline
\end{tabular}
\label{tab:para}
\end{center}
\end{table*}

\subsection{quark matter}
\label{sec:NJL}

We adopt the three flavor NJL model to describe the deconfined quark phase.
The Lagrangian is given by
\begin{eqnarray}
\label{eq:Lnjl}
\mathcal{L}_{\rm{NJL}} &=&\bar{q}\left( i\gamma _{\mu }\partial ^{\mu
}-m^{0}\right) q+{G_S}\sum\limits_{a = 0}^8 {\left[ {{{\left( {\bar q{\lambda _a}q} \right)}^2}
+ {{\left( {\bar q i{\gamma _5}{\lambda _a}q} \right)}^2}} \right]}  \nonumber \\
&&-K\left\{ \det \left[ \bar{q}\left( 1+\gamma _{5}\right) q\right] +\det %
\left[ \bar{q}\left( 1-\gamma _{5}\right) q\right] \right\} +\mathcal{L}_{V}
 , \nonumber \\
\end{eqnarray}%
with
\begin{align}
{{\cal L}_V}
=&  - {G_0}{\left( {\bar q{\gamma ^\mu }q} \right)^2}
    - {G_V}\sum_{\alpha=1}^8
      \left[ \left(\bar{q} \gamma ^\mu        \lambda_\alpha q\right)^2
           + \left(\bar{q}i\gamma ^\mu\gamma_5\lambda_\alpha q\right)^2
	\right]
\ ,
\end{align}
in which $q$ denotes the quark field with three flavors ($N_f$ =3) and
three colors ($N_c$ =3).
The determinant interaction is included
in order to take account of the $\mathrm{U}(1)_A$ anomaly.
$G_S$, $G_0$, and $G_V$ are the scalar,
flavor-singlet-vector, and flavor-octet-vector coupling constants, respectively,
and have dimensions of energy$^{-2}$.
In the mean field approximation, only those terms with diagonal matrix elements
in $\lambda_\alpha$ remain, then ${\cal L}_V$ is reduced to 
\begin{eqnarray}
{{\cal L}_V} &=&  - {G_0}{\left( {\bar q{\gamma ^\mu }q} \right)^2} 
- {G_3}\left[ {{{\left( {\bar q{\gamma ^\mu }{\lambda _3}q} \right)}^2} 
+ {{\left( {\bar qi{\gamma ^\mu }{\gamma _5}{\lambda _3}q} \right)}^2}} \right] \nonumber \\
&&- {G_8}\left[ {{{\left( {\bar q{\gamma ^\mu }{\lambda _8}q} \right)}^2} 
+ {{\left( {\bar qi{\gamma ^\mu }{\gamma _5}{\lambda _8}q} \right)}^2}} \right] .
\end{eqnarray}
In the flavor SU(3) limit, 
the isovector-vector coupling ($G_3$) and hypercharge-vector coupling ($G_8$) constants
should be the same, $G_3=G_8=G_V$.
In order to discuss the (isospin) symmetry energy
and the hypercharge symmetry energy effects separately,
we consider the cases with $G_3 \not= G_8$ as well.

In the mean field approximation, quarks get constituent quark masses by spontaneous
chiral symmetry breaking,
\begin{equation}
\label{eq:gap}
m_{i}^{\ast }=m_{i}^{0}-4{G_S}\langle \bar{q}_{i}q_{i}\rangle +2K\langle \bar{q}%
_{j}q_{j}\rangle \langle \bar{q}_{k}q_{k}\rangle,
\end{equation}%
where $\left\langle \bar{q}_{i}q_{i}\right\rangle \equiv C_i$
denotes the quark scalar density,
and $(i,j,k)$ is a permutation of $(u, d, s)$.
For charge neutral quark matter containing quarks ($u$, $d$, and $s$) and
leptons ($e$ and $\mu$) in $\beta $ equilibrium,
the chemical potentials satisfy the relations
$\mu_{s}=\mu_{d}=\mu_{u}+\mu_{e}$ and $\mu_{\mu}=\mu_{e}$,
where the chemical potentials of $u$, $d$, and $s$ quarks are given by
\begin{eqnarray}
{\mu _u} &=& \frac{{\partial \varepsilon }}{{\partial {n_u}}} = \sqrt {k{{_F^u}^2} + m_u^{*2}}  
+ 2{G_0}\left( {{n_u} + {n_d} + {n_s}} \right) \nonumber\\
&&+ 2{G_3}\left( {{n_u} - {n_d}} \right) + \frac{2}{3}{G_8}\left( {{n_u} + {n_d} - 2{n_s}} \right), \\
 {\mu _d} &=& \frac{{\partial \varepsilon }}{{\partial {n_d}}} = \sqrt {k{{_F^d}^2} + m_d^{*2}} 
  + 2{G_0}\left( {{n_u} + {n_d} + {n_s}} \right) \nonumber\\
 &&- 2{G_3}\left( {{n_u} - {n_d}} \right) + \frac{2}{3}{G_8}\left( {{n_u} + {n_d} - 2{n_s}} \right), \\
 {\mu _s} &=& \frac{{\partial \varepsilon }}{{\partial {n_s}}} = \sqrt {k{{_F^s}^2} + m_s^{*2}}  
 + 2{G_0}\left( {{n_u} + {n_d} + {n_s}} \right) \nonumber\\
 &&- \frac{4}{3}{G_8}\left( {{n_u} + {n_d} - 2{n_s}} \right).\label{eq:mus}
\end{eqnarray}%
The energy density of quark matter is given by
\begin{eqnarray}
\label{eq:eNJL}
\varepsilon_{\rm{NJL}} &=&\sum\limits_{i = u,d,s}
 {\left[ { - \frac{3}{{{\pi ^2}}}\int_{k_F^i}^\Lambda
  {\sqrt {{k^2} + m_i^{ * 2}} } \;{k^2}dk} \right]}
\nonumber\\
&&   + 2{G_S}\left( {C_u^2 + C_d^2 + C_s^2} \right)
- 4K{C_u}{C_d}{C_s}
\notag\\
&& + {G_0}{\left( {{n_u} + {n_d} + {n_s}} \right)^2} \notag\\
&&+ {G_3}{\left( {{n_u} - {n_d}} \right)^2} \notag\\
&&+ \frac{1}{3}{G_8}{\left( {{n_u} + {n_d} - 2{n_s}} \right)^2}   - {\varepsilon _0},
\end{eqnarray}%
where $\varepsilon_{0}$ is subtracted to ensure $\varepsilon_{\rm{NJL}}=0$ in
the vacuum. The total energy density and pressure for quark matter are given by
\begin{eqnarray}
\label{eq:e2}
\varepsilon_{\rm{QP}} &=& \varepsilon_{\rm{NJL}}
  +\varepsilon_{l},
\\
P_{\rm{QP}} &=&\sum_{i=u,d,s,e,\mu }n_{i}\mu_{i}-\varepsilon_{\rm{QP}}.
\label{eq:p2}
\end{eqnarray}

We employ the parameter set
given in Ref.~\cite{Rehb96}, $m_{u}^{0}=m_{d}^{0}=5.5\ \text{MeV}$,
$m_{s}^{0}=140.7\ \text{MeV}$, $\Lambda =602.3\ \text{MeV}$, ${G_S}\Lambda^{2}=1.835$,
and $K\Lambda ^{5}=12.36$.
The vector couplings ($G_0, G_3, G_8$) are considered as free parameters,
and we use $G_0=0.25\,G_S$, $G_3,G_8=(0,\,1.5,\,10)\ G_0$.
For larger $G_0$ values ($G_0 > 0.27\,G_S$),
the energy in quark matter is found to be always larger
than that in hadronic matter at high densities,
and there is no phase transition.
The parameter choice of $G_3=G_8=1.5\,G_0$ corresponds to
the Lagrangian adopted in Refs.~\cite{Wu17,Rehb96}.
The larger isovector-vector coupling, $G_3=G_8=10\,G_0$,
roughly gives the symmetry energy slope of $L_Q\simeq 50~\MeV$.
The vector couplings increase the energy per baryon as
\begin{align}
\frac{\Delta\varepsilon_V}{n_b}
=&9\,G_0\,n_b
+G_3\,n_b\,\delta^2
+3\,G_8\,n_b\,\delta_h^2
\ ,\\
\delta=& \frac{n_d-n_u}{n_b}
\ ,\quad
\delta_h= \frac{n_b-n_s}{n_b}=\frac{B+S}{B}
\ .
\end{align}
There are two types of asymmetry parameter,
$\delta$ and $\delta_h$,
which are the isospin asymmetry
and the hypercharge ($Y=B+S$) fraction.
The symmetry energy is defined as the coefficient of $\delta^2$,
then the vector coupling contribution to the symmetry energy is given as
\begin{align}
\Delta S_V(n_b)=&G_3\,n_b
=G_3\,n_0 + 3\,G_3\,n_0\,\left(\frac{n_b-n_0}{3n_0}\right)
\ .
\end{align}
The vector coupling contribution to the slope parameter
is $\Delta L_V=3G_3n_0=6.6$ and 44 MeV
for $G_3=1.5G_0$ and $10G_0$, respectively.

\begin{figure}[htp]
\begin{minipage}[t]{1\linewidth}
\begin{center}
\includegraphics[width=7 cm,clip]{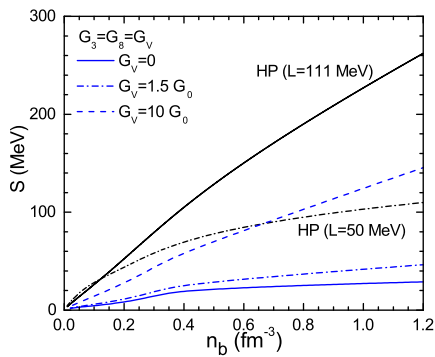}
\end{center}
\caption{(Color online) Symmetry energy as a function of the baryon number density
for the pure hadronic phase (black lines) and the pure quark phase (blue lines) with different vector couplings.}
\label{fig:1nbesym}
  \end{minipage}
\end{figure}

In Fig.~\ref{fig:1nbesym}, we compare the symmetry energy $S$ as a function
of  baryon number density $n_b$ for the pure hadronic matter and pure quark matter.
The results of $G_3=G_8=0, 1.5G_0$, and $10G_0$ are shown. Usually the quark matter
symmetry  energy is much smaller than the TM1e with $G_3=G_8=0$ or $1.5G_0$, 
while with $G_3=G_8=10G_0$, the result is comparable to TM1e ($L=50$ MeV).
Thus the parameter sets with $G_3=G_8=(0-10)G_0$ covers a wide and reasonable
range of symmetry energy.

\subsection{Hadron-quark phase transition}
\label{sec:Gibbs}
For neutron-star matter, the $\beta$ equilibrium and
the charge neutrality are satisfied.
We adopt simple Gibbs conditions for the mixed phase
connecting the pure hadronic phase and the pure quark phase.
The mixed phase may appear in the inner core of the neutron stars.

In the Gibbs conditions, the global charge neutrality condition is given by
\begin{equation}
un_c^{{\rm{QP}}} + \left( {1 - u} \right)n_c^{{\rm{HP}}} = 0
,  \label{eq:charge3}
\end{equation}%
in which $u=V_{\rm{QP}}/(V_{\rm{QP}}+V_{\rm{HP}})$
represents the volume fraction of quark matter in the mixed phase.
The mechanical equilibrium requires
\begin{equation}
{P_{{\rm{HP}}}}\left( {{\mu _n},{\mu _e}} \right) = {P_{{\rm{QP}}}}\left( {{\mu _n},{\mu _e}} \right).
\label{eq:p3}
\end{equation}%
There are two independent chemical potentials, $\mu_n$ and $\mu_e$; the hadronic and quark phases
satisfy the chemical equilibrium condition,
\begin{equation}
\mu_u+\mu_e=\mu_d=\mu_s=\frac{\mu_n}{3}+\frac{\mu_e}{3}.
\end{equation}%
With these equilibrium constraints, we can solve the mixed phase self-consistently and obtain the
properties of the hadron-quark mixed phase.

\section{Isovector-vector coupling dependence of hadron-quark coexistence
in neutron-star matter}
\label{sec:results}

\begin{table*}[bthp]
\caption{Model dependence of the coexistence densities
and neutron-star properties, $G_0\,=\,0.25G_S$.}
\begin{center}
\begin{tabular}{lcccccccccccc}
\hline\hline
Model  &$L$    &$G_3/G_0$ &$G_8/G_0$  &$n_b^{(1)}$  &$n_b^{(2)}$  &${M_{\rm{max}}}/{M_\odot}$  &$n_c^{M_{\rm{max}}}$
        &$R(1.4\ M_\odot)$ \\
        & (MeV) & & &$(\rm{fm^{-3}})$ &  $(\rm{fm^{-3}})$  & &$(\rm{fm^{-3}})$ &  (km)\\
\hline
TM1               & 110.8   & -  &-      &-        &-           & 2.180 & 0.871  &14.3       \\
TM1/NJL-V    & 110.8   &0   &0    &0.508  &2.209  &2.098  &0.769  &14.3  \\
TM1/NJL-VR1  & 110.8   &1.5 &0    &0.565  &2.168  &2.125  &0.816  &14.3  \\
TM1/NJL-VRY1 & 110.8   &1.5 &1.5  &0.565  &2.225  &2.125  &0.815  &14.3  \\
TM1/NJL-VR2  & 110.8   &10  &0    &0.687  &2.067  &2.163  &0.850  &14.3  \\
TM1/NJL-VRY2 & 110.8   &10  &10   &0.718  &2.240  &2.170  &0.830  &14.3  \\
\hline
TM1e               & 50   & -  &-      &-        &-           & 2.122 & 0.899  &13.0       \\
TM1e/NJL-V    & 50     &0    &0    &0.681  &2.210  &2.103  &0.900  &13.0  \\
TM1e/NJL-VR1  & 50     &1.5  &0    &0.753  &2.170  &2.114  &0.881  &13.0  \\
TM1e/NJL-VRY1 & 50     &1.5  &1.5  &0.757  &2.226  &2.115  &0.879  &13.0  \\
TM1e/NJL-VR2  & 50     &10   &0    &0.890  &2.076  &2.122  &0.888  &13.0  \\
TM1e/NJL-VRY2 & 50     &10   &10   &0.938  &2.241  &2.122  &0.888  &13.0  \\
\hline
 &     &  &  &  &  & 1.928 $\pm$ 0.017 ~\cite{Demo10,Fons16}    &  &12 $\pm$ 1 ~\cite{Steiner16} \\
Constraints & 40$\sim$60 ~\cite{Tews17} &-  &-  &-  &-      & 2.01 $\pm$ 0.04 ~\cite{Anto13} &-  &9.4 $\pm$ 1.2 ~\cite{Guillot14}  \\
&     &  &  &   &   &  &  &$>14$~\cite{Haensel:2016th} \\
\hline\hline
\end{tabular}
\label{tab:2coex}
\end{center}
\end{table*}

We shall now investigate the effects of isovector-vector coupling on the equation
of state, the density range of hadron-quark coexistence,
and the properties of neutron stars.
For this intent, we use the RMF and NJL models
to describe hadronic and quark matter, respectively.
For RMF, the TM1 and TM1e parameter sets are used.
These parameter sets show different symmetry energy slopes ($L$),
which show significant effects on the neutron-star radius. We include
isovector-vector coupling in NJL, which
modifies the quark-matter symmetry energy.
In the following discussions,
we fix the isoscalar-vector coupling as $G_0=0.25G_S$,
and we compare the results of
$(G_3/G_0,G_8/G_0)=(0,0),(1.5,0),(1.5,1.5),(10,0)$, and $(10,10)$,
referred to as 
NJL-V, NJL-VR1, NJL-VRY1, NJL-VR2, and NJL-VRY2, respectively.
NJL-V and NJL-VRY1 corresponds to models
in Ref.~\cite{Benic:2014jia} and Refs.~\cite{Wu17,Rehb96}, respectively.

We define $n_b^{(1)}$ and $n_b^{(2)}$ as the starting
and the ending baryon densities of the mixed phase.
At the density $n_b^{(1)}$,
the energy per baryon in the mixed phase becomes lower than that of pure hadronic phase.
The volume fraction of quark matter $u$ increases with the baryon number density $n_b$,
and it transforms into pure quark phase at the density $n_b^{(2)}$
under the condition that the mixed phase has larger energy density than the pure quark phase.
The model dependence of the phase transition densities $n_b^{(1)}$ and
$n_b^{(2)}$ is summarized in Table~\ref{tab:2coex}.

\subsection{Equation of state}

\begin{figure*}[tbhp]
\includegraphics[bb=5 5 580 420, width=5 cm,clip]{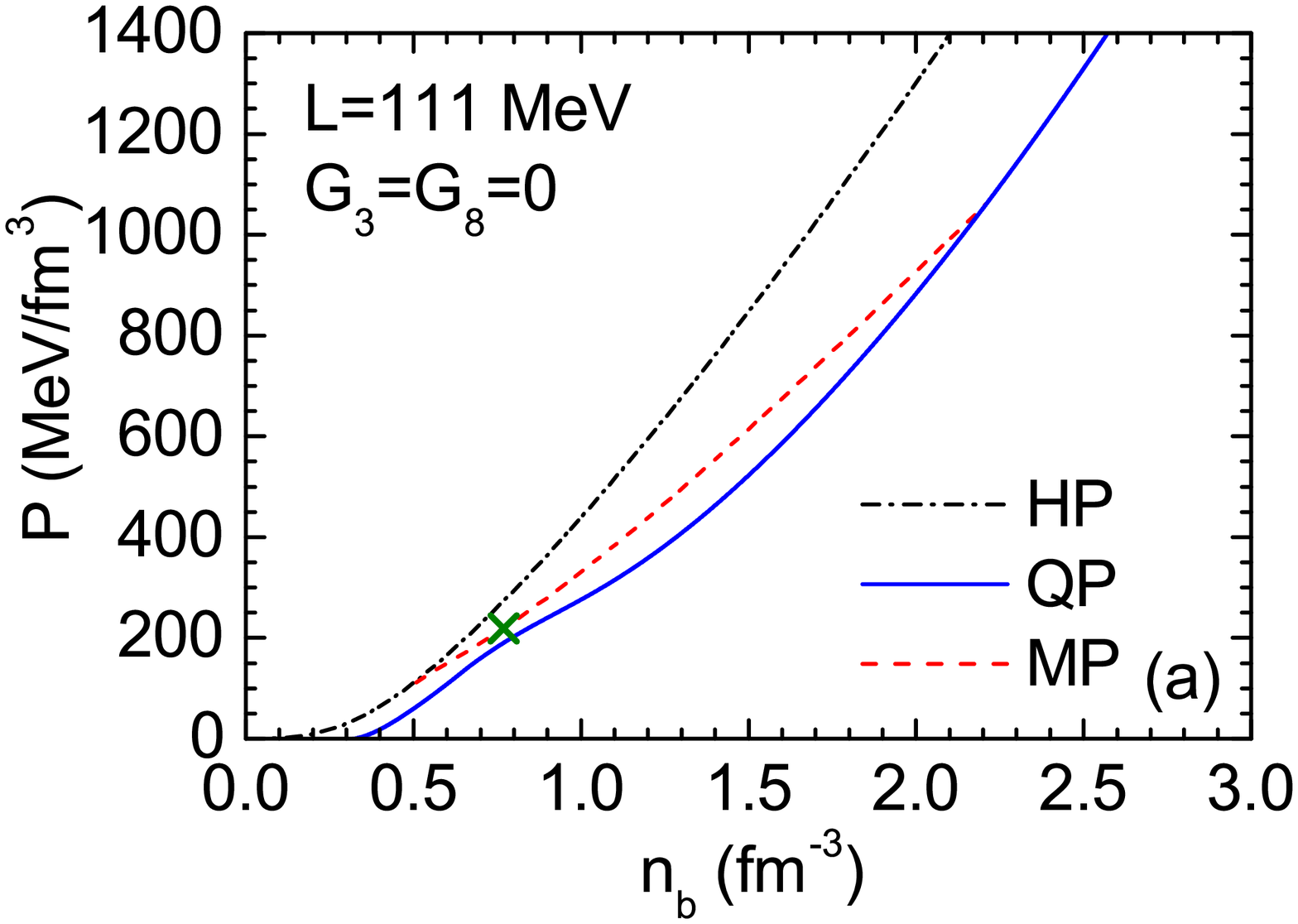}
\includegraphics[bb=5 5 580 420, width=5 cm,clip]{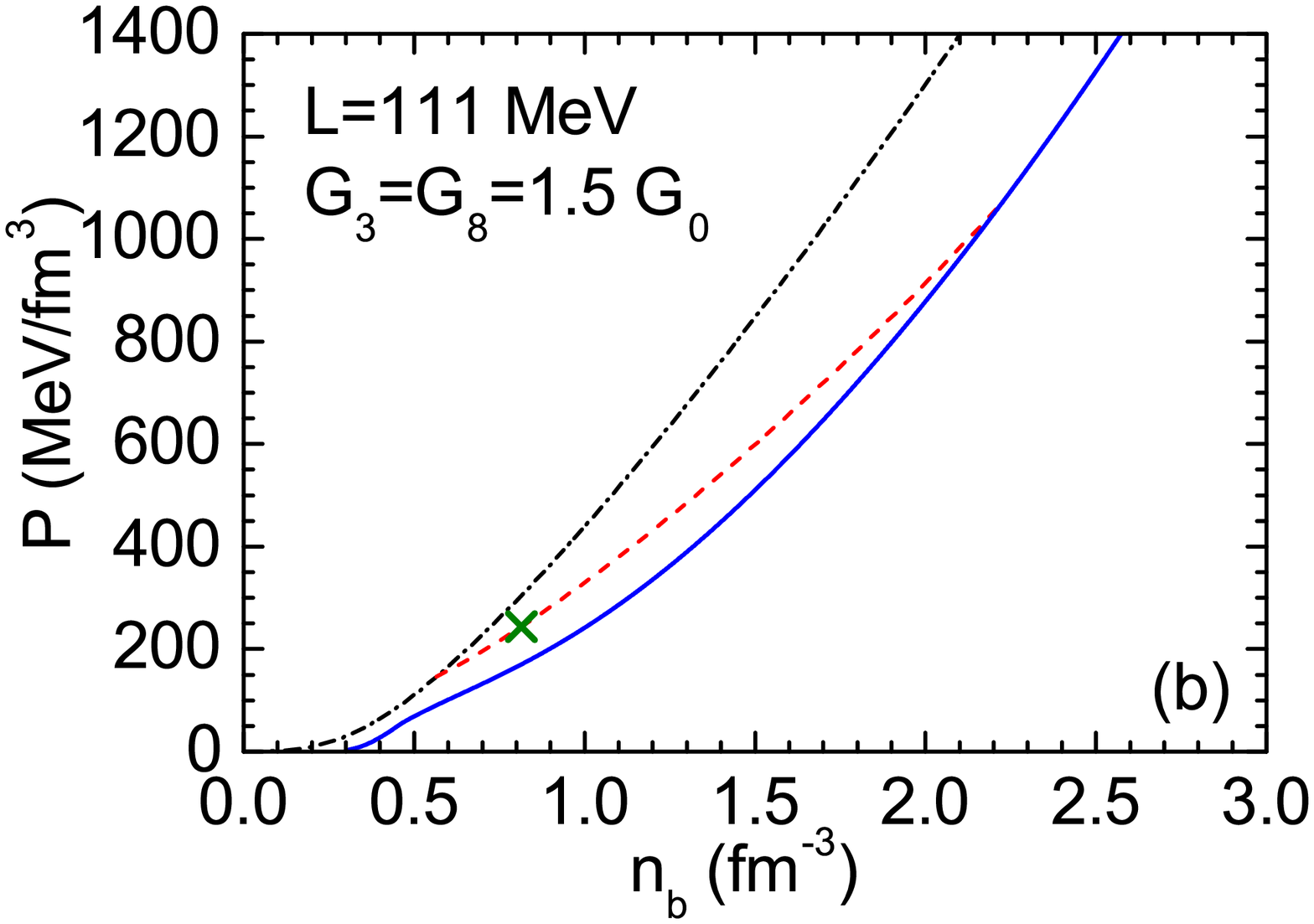}
\includegraphics[bb=5 5 580 420, width=5 cm,clip]{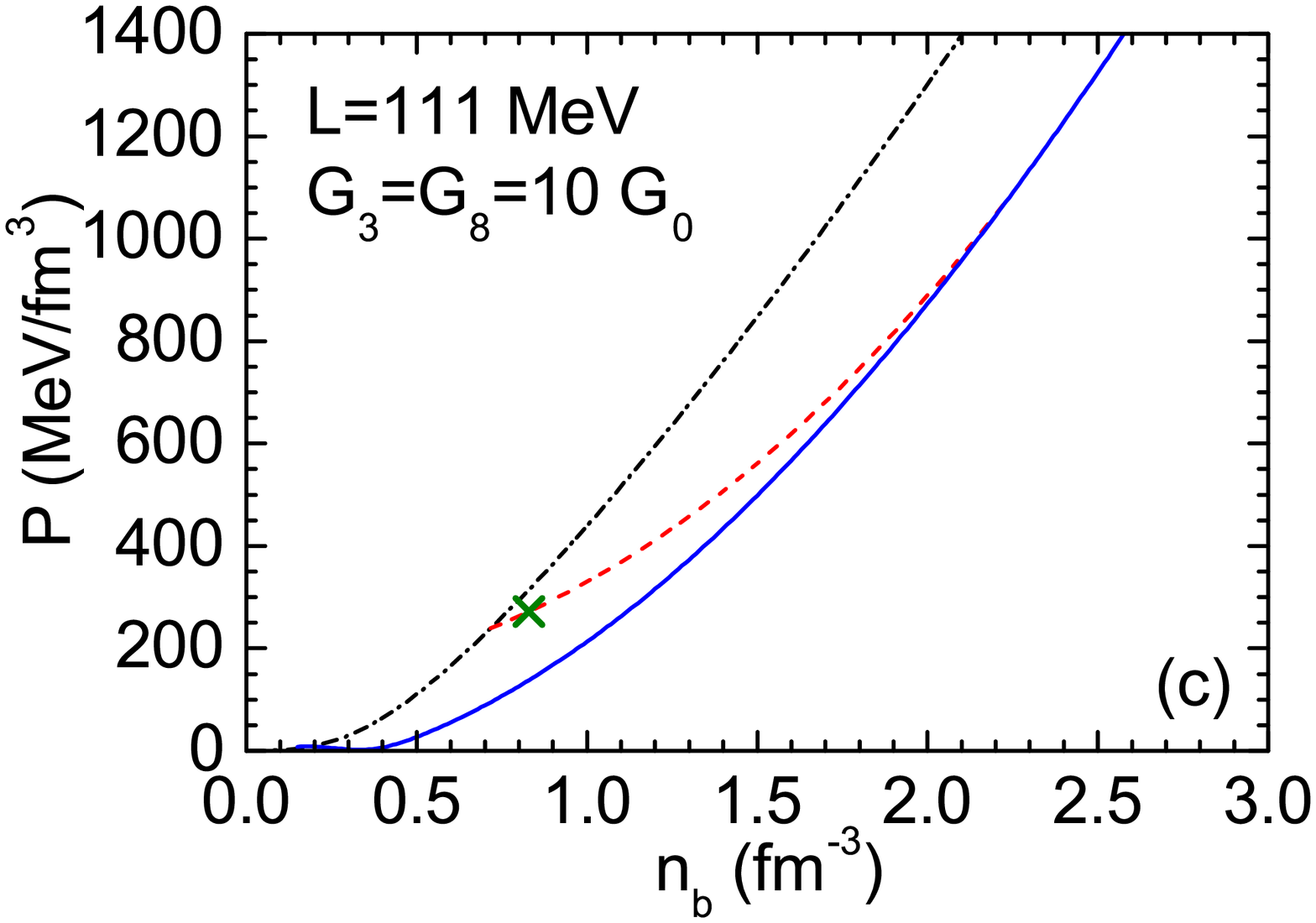}\\
\includegraphics[bb=0 0 580 420, width=5 cm,clip]{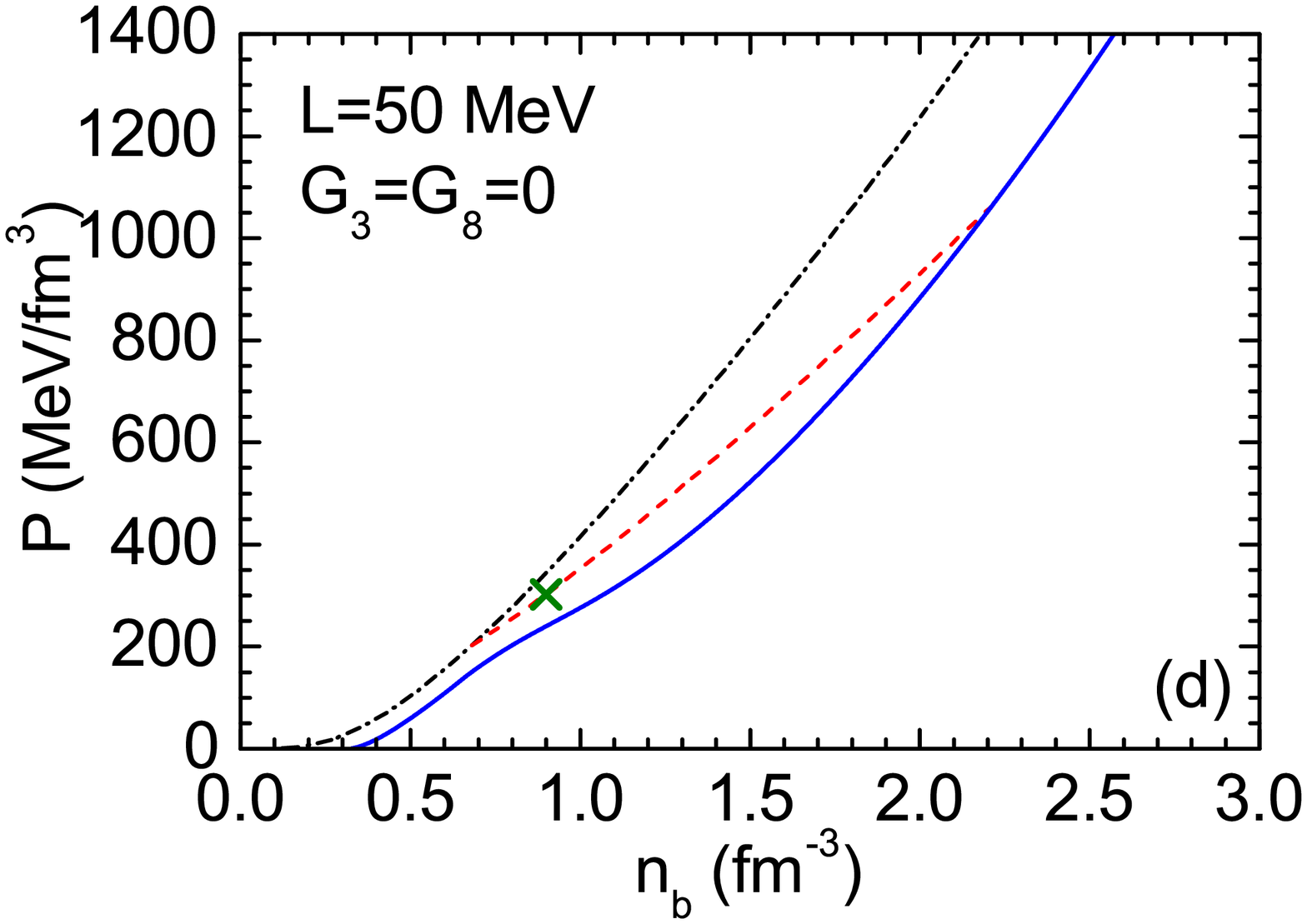}
\includegraphics[bb=0 0 580 420, width=5 cm,clip]{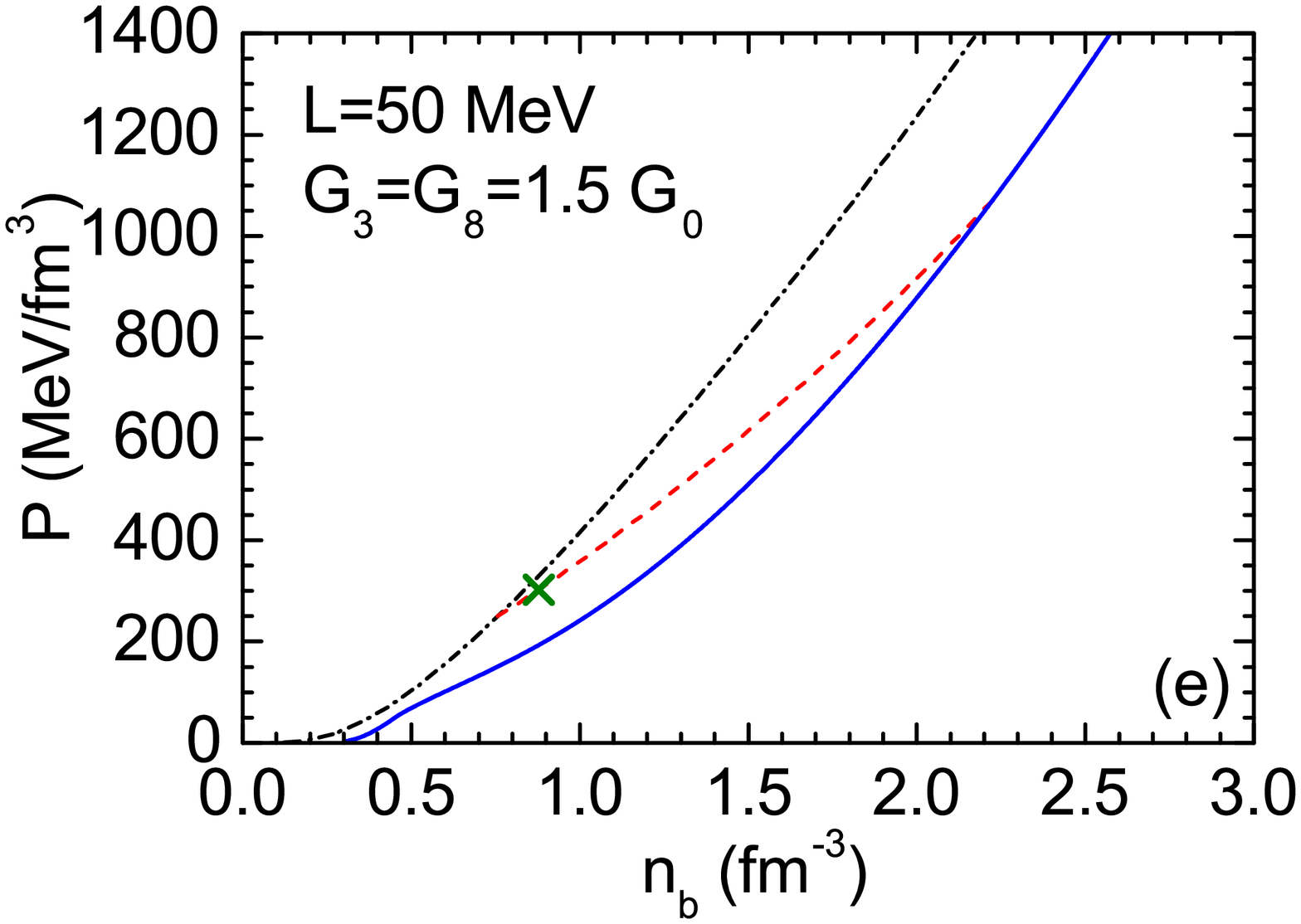}
\includegraphics[bb=5 5 580 420, width=5 cm,clip]{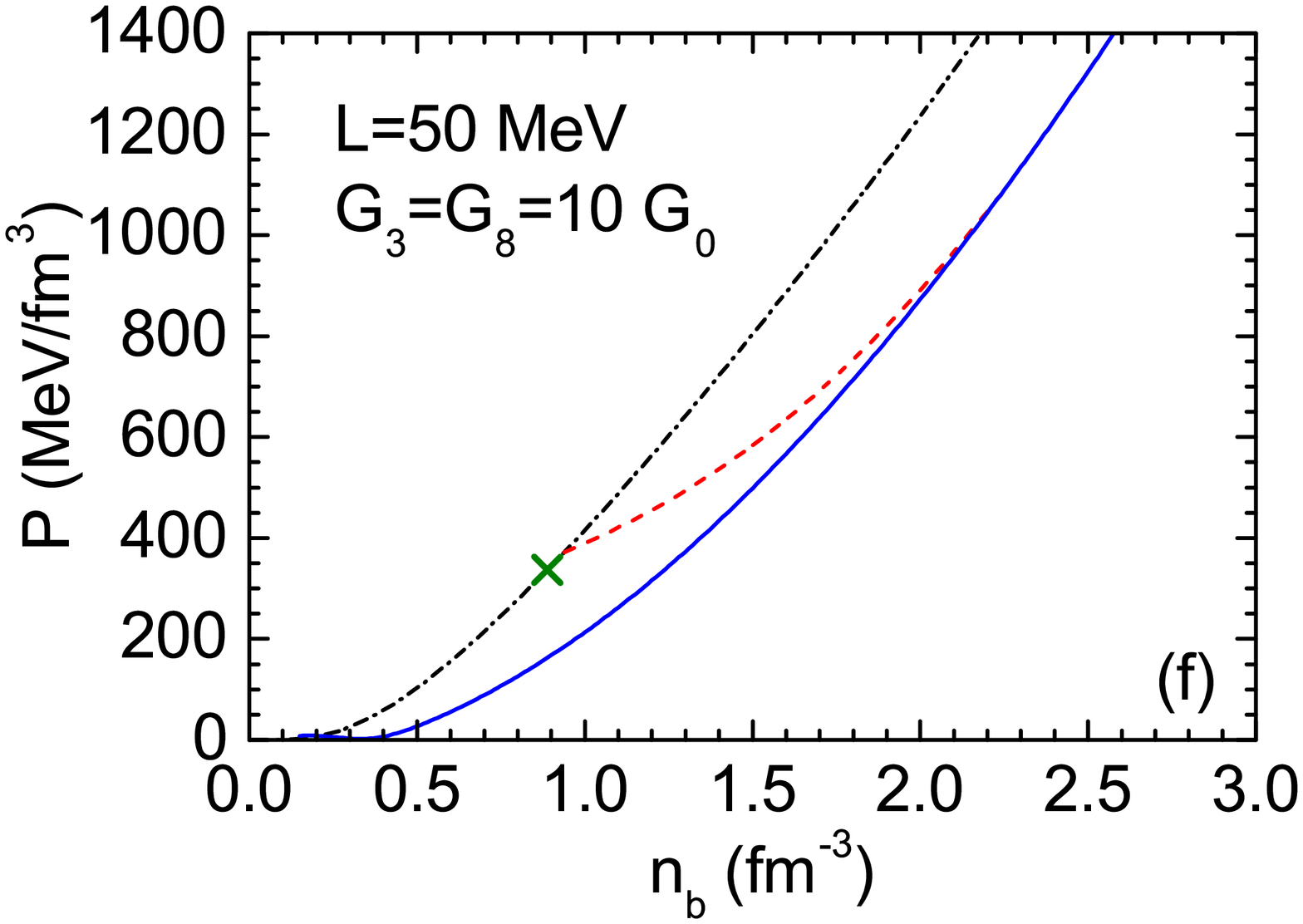}
\caption{(Color online) Pressures as a function of the baryon number density
for the pure hadronic phase (dash-dotted lines), the mixed phase (dashed lines), 
and the pure quark phase (solid lines).}
\label{fig:2nbp}
\end{figure*}

We first discuss the equation of state.
In Fig.~\ref{fig:2nbp}, we show the pressure
as a function of the baryon number density for hadronic matter, quark matter,
and the mixed phase obtained under Gibbs conditions.
The top and bottom panels show the results from the TM1
($L=110.8~\MeV$) and TM1e ($L=50~\MeV$) parameter sets, respectively.
The left, middle, and right panels show the results of
NJL-V ($G_3=G_8=0$),
NJL-VRY1 ($G_3=G_8=1.5G_0$),
and NJL-VRY2 ($G_3=G_8=10G_0$), respectively.

\begin{figure*}[htbp]
\includegraphics[bb=40 5 580 580, width=7 cm,clip]{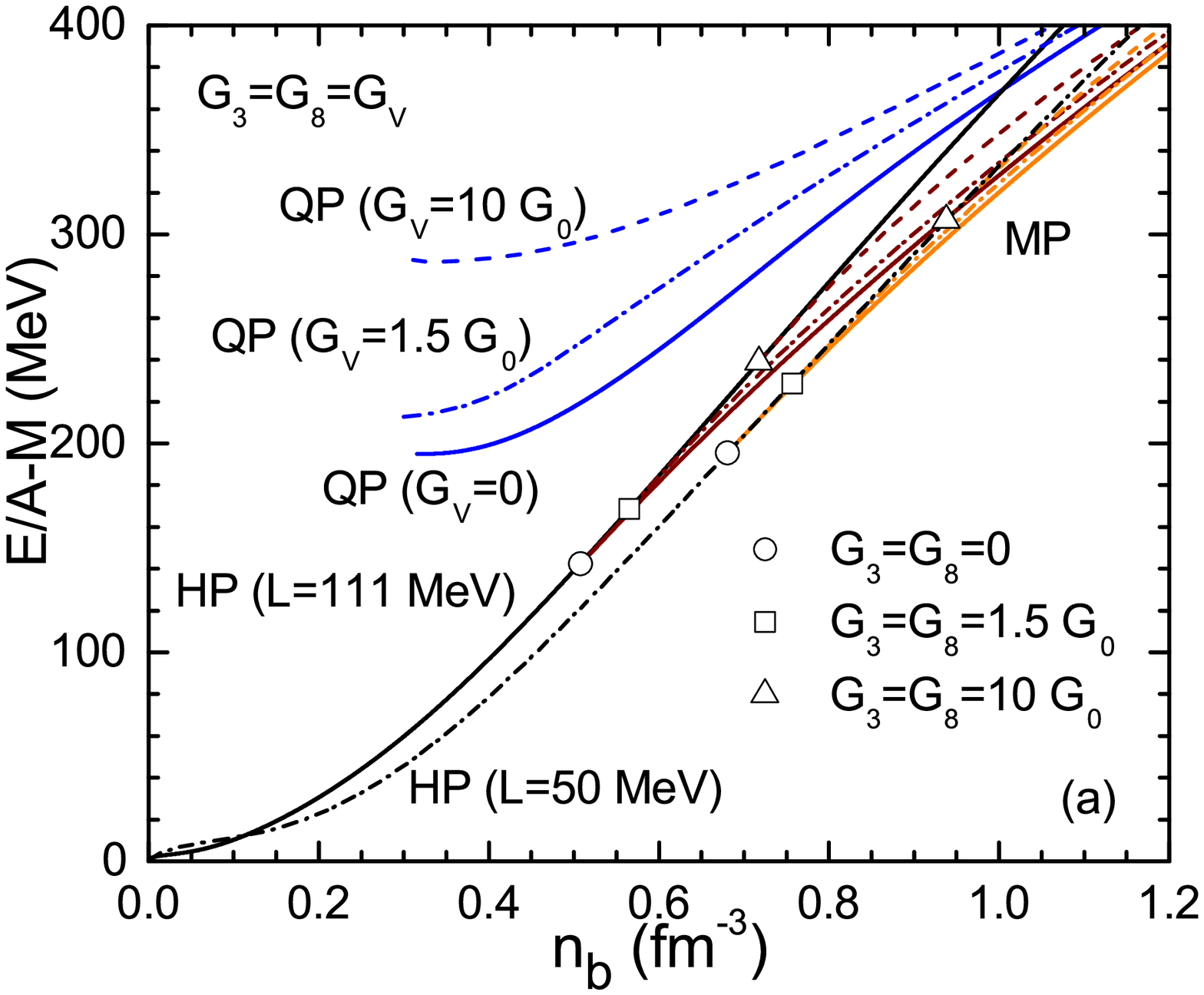}%
\includegraphics[bb=40 5 580 580, width=7 cm,clip]{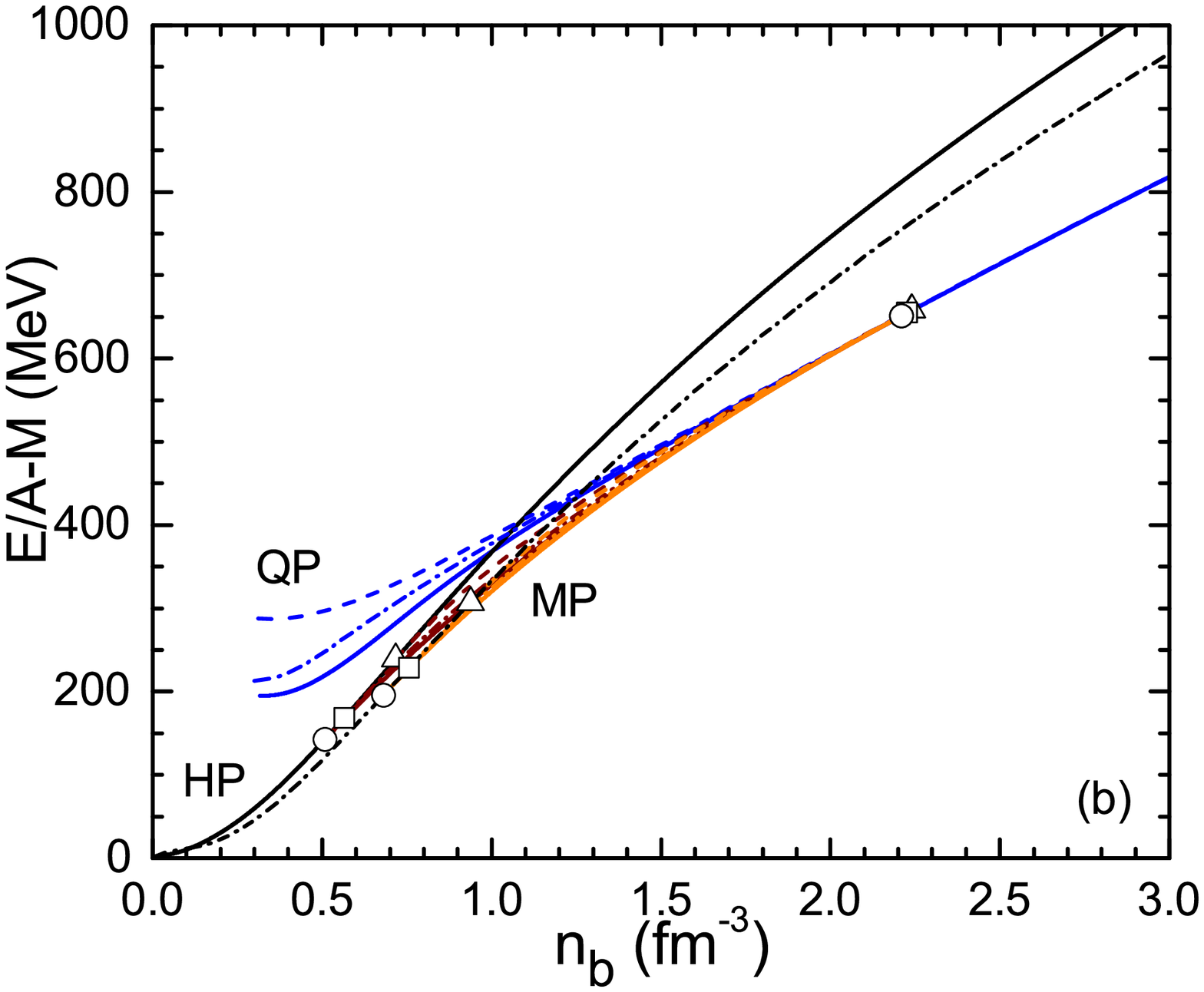}
\caption{(Color online) Energy density per baryon as a function of the
baryon number density for the hadronic, mixed, and quark phases. The left panel is
a local enlargement of the right panel. Open circles (squares, triangles) label the transition
points for $G_3=G_8=0$ ($G_3=G_8=1.5G_0$, $G_3=G_8=10G_0$).}
\label{fig:3nbea}
\end{figure*}

In Fig.~\ref{fig:3nbea}, we compare the energy per baryon , $E/A-M$,
as a function of baryon number density $n_b$ for the hadronic, mixed, and quark phases.
Open circles (squares, triangles) show the transition densities of NJL-V  (NJL-VRY1, NJL-VRY2).
We find isovector-vector and hypercharge-vector couplings delay the transition to the mixed phase.
Although the energy difference between NJL-V, NJL-VR1, and NJL-VR2 is
small in quark matter,  the $n_b^{(1)}$ difference is very visible,
whereas the $n_b^{(2)}$ difference is rather small.
The reason comes from the difference of the isospin and hypercharge asymmetry of quarks,
$\delta_\mathrm{Q}=3(n_d-n_u)/(n_d+n_u+n_s)$ and $\delta_{h}^\mathrm{Q}=(n_u+n_d-2n_s)/(n_d+n_u+n_s)$,
in the pure quark phase and in the mixed phase.
In the mixed phase, the quark matter part is negatively
charged and the number density difference between $u$ quark and $d$ quark (so as hypercharge difference)
is bigger than that in the charge neutral pure quark phase as can be found from the electron chemical 
potential discussed below.

\begin{figure*}[htbp]
\includegraphics[bb=40 5 580 580, width=7 cm,clip]{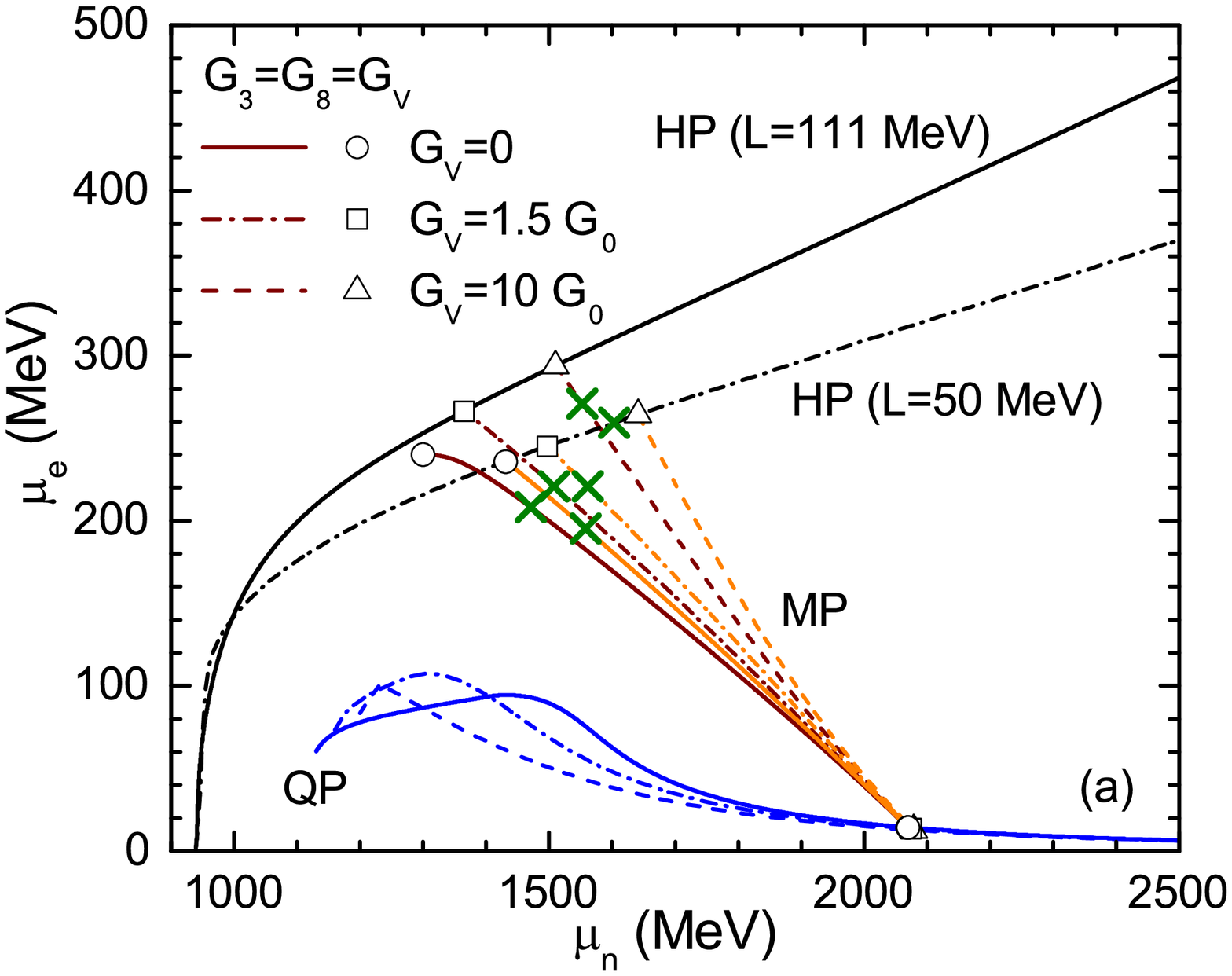}%
\includegraphics[bb=40 5 580 580, width=7 cm,clip]{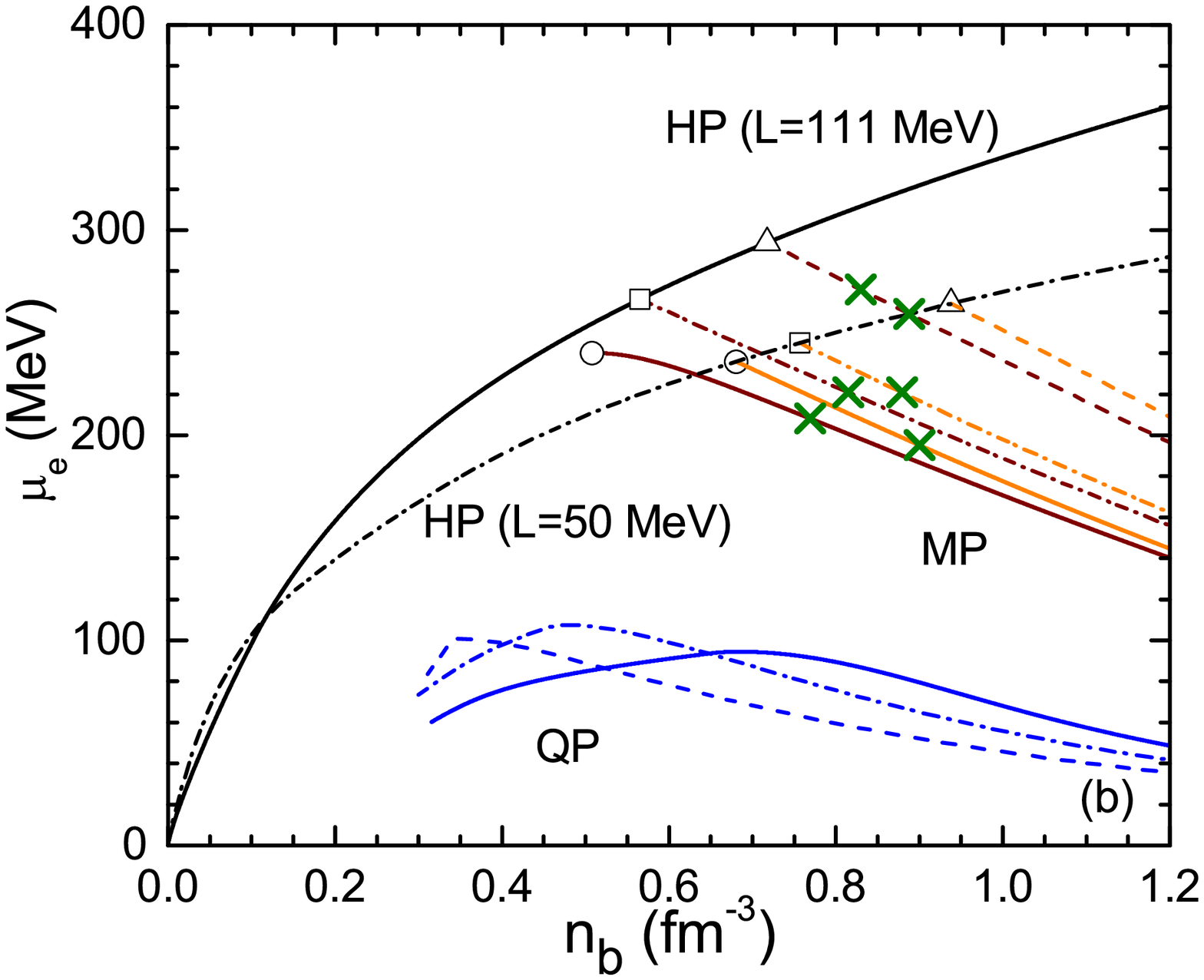}
\caption{(Color online) The left panel shows the relation between the chemical
 potential of neutron and electron. The right panel is the electron chemical
 potential as a function of baryon number density for
 different phases.}
\label{fig:4xmunxmue}
\end{figure*}

We show the electron chemical potential
as a function of the neutron chemical potential (the baryon density) in
the left (right) panel of Fig.~\ref{fig:4xmunxmue}.
The electron chemical potential $\mu_e$ in the mixed phase is significantly
larger than that in the pure quark matter at the same baryon number density $n_b$.
The electron chemical potential $\mu_e$ reflects the chemical potential
difference of the $u$ and $d$ quarks ($\mu_d-\mu_u=\mu_e$). It can be seen as a signal 
of the imbalance between $u$ and $d$ quarks. 
The system becomes more symmetric with decreasing $\mu_e$.
The behavior of $\mu_e$ can explain why the effect of isovector-vector coupling is more
significant at lower densities.
For pure quark matter, there exists a maximum value of $\mu_e$,
which corresponds to the appearance of $s$ quark. This change expresses that the
imbalance between $u$ and $d$ quarks is getting smaller. These trends are the same as those found in
Fig.~\ref{fig:3nbea}. 

It would be interesting to discuss the reason why the electron chemical
potential is small in pure quark matter.
In Fig.~\ref{fig:4xmunxmue},
we find that $\mu_e$ in pure quark matter increases at low densities,
reaches $\mu_e \simeq 100~\MeV$,
and turns to decrease at around $\mu_n=1300~\MeV$.
At this density, the quark chemical potentials are evaluated as
$\mu_u \simeq 370~\MeV$ and $\mu_d=\mu_s \simeq 470~\MeV$.
Since the chemical potential of $s$ quark is close to its threshold value
for appearance,
then we expect that the appearance of $s$ quarks would be the mechanism
to suppress the electron chemical potential.

We find that the mixed phase shrinks when the isovector-vector coupling ($G_3$)
and hypercharge-vector coupling ($G_8$) are switched on.
Two couplings, $G_3$ and $G_8$, 
modify both of the transition densities, and the shift of $n_b^{(1)}$ is larger
than that of $n_b^{(2)}$.
This difference comes from the density dependence of the isospin asymmetry,
$\delta\equiv ((1-u)(n_n-n_p)+u(n_d-n_u))/n_b$.
In the hadron-quark mixed phase, the matter tends to be more symmetric with 
increasing baryon number density $n_b$.
At $n_b \simeq n_b^{(1)}< 0.9~\fm^{-3}$, the isospin asymmetry $\delta$ is still significant,
while $\delta$ is almost zero at $n_b \simeq n_b^{(2)} > 2.3~\fm^{-3}$.
Since the energy from the isovector-vector coupling is proportional to
$\delta^2$, it becomes small at high baryon densities. This mechanism also applies to
the effect of the symmetry energy slope $L$.

\subsection{Particle fraction and asymmetry}

\begin{figure*}[tbp]
\includegraphics[bb=40 5 580 400, width=7 cm,clip]{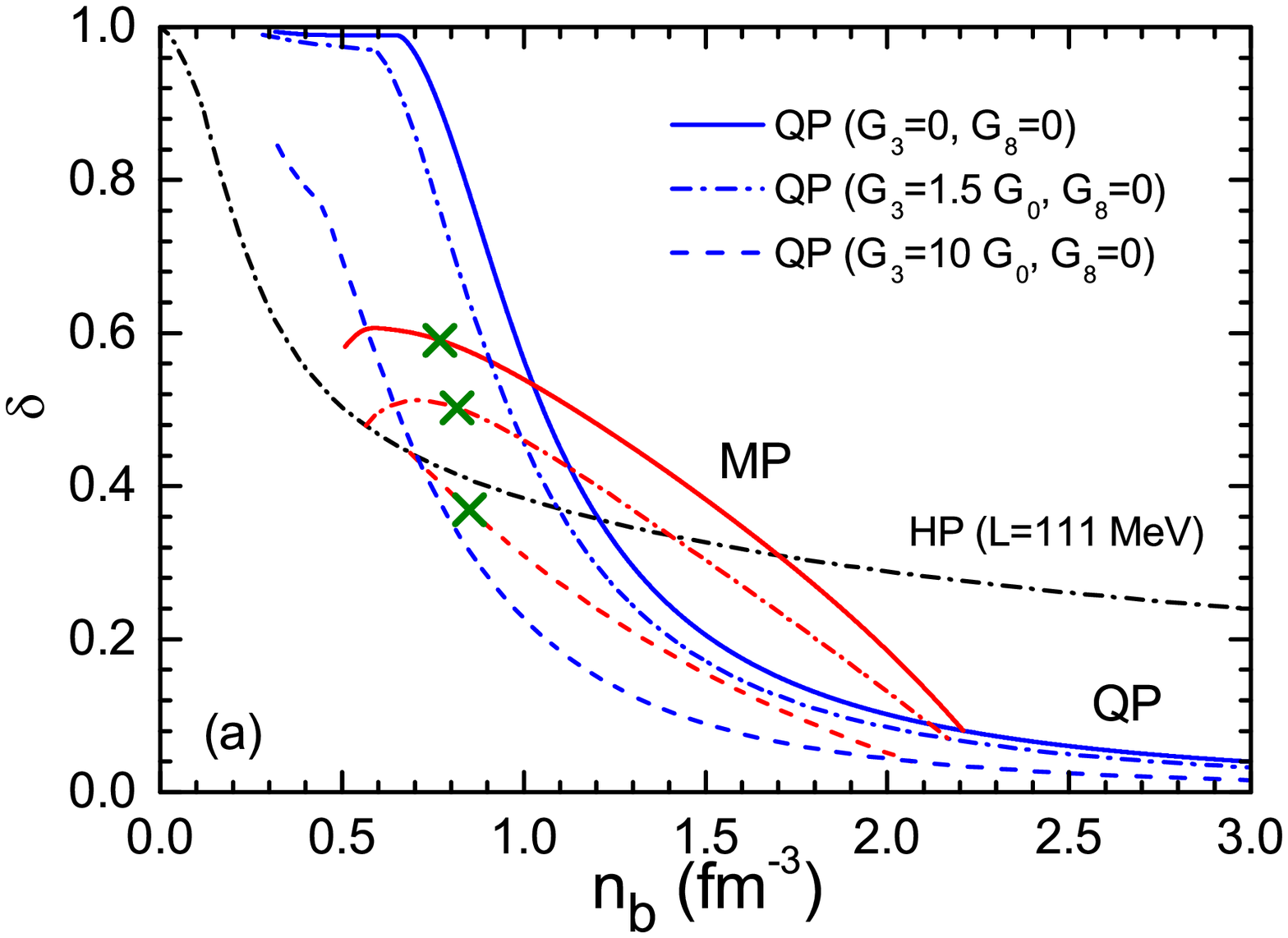}%
\includegraphics[bb=40 5 580 400, width=7 cm,clip]{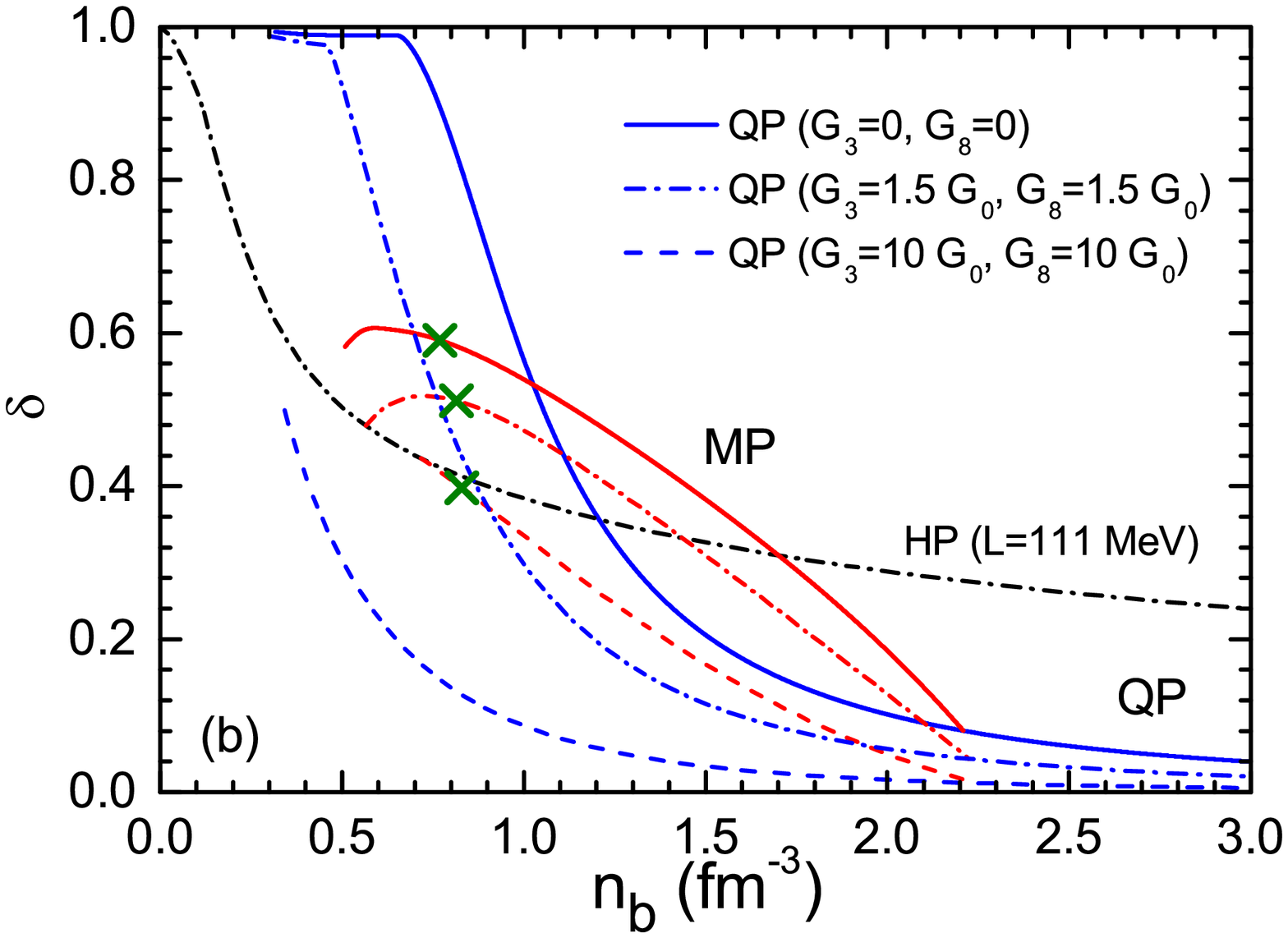} \\
\includegraphics[bb=40 5 580 400, width=7 cm,clip]{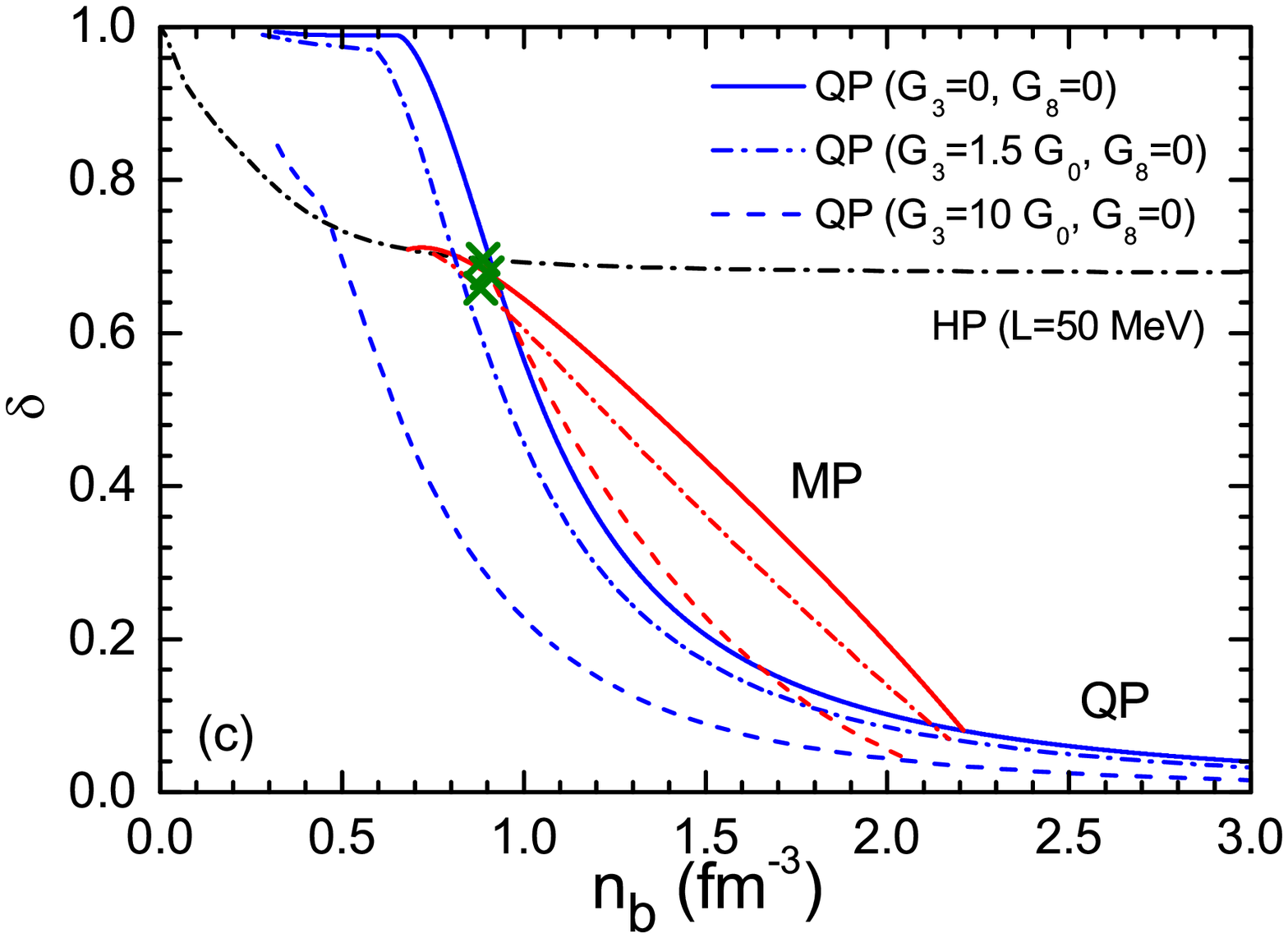}%
\includegraphics[bb=40 5 580 400, width=7 cm,clip]{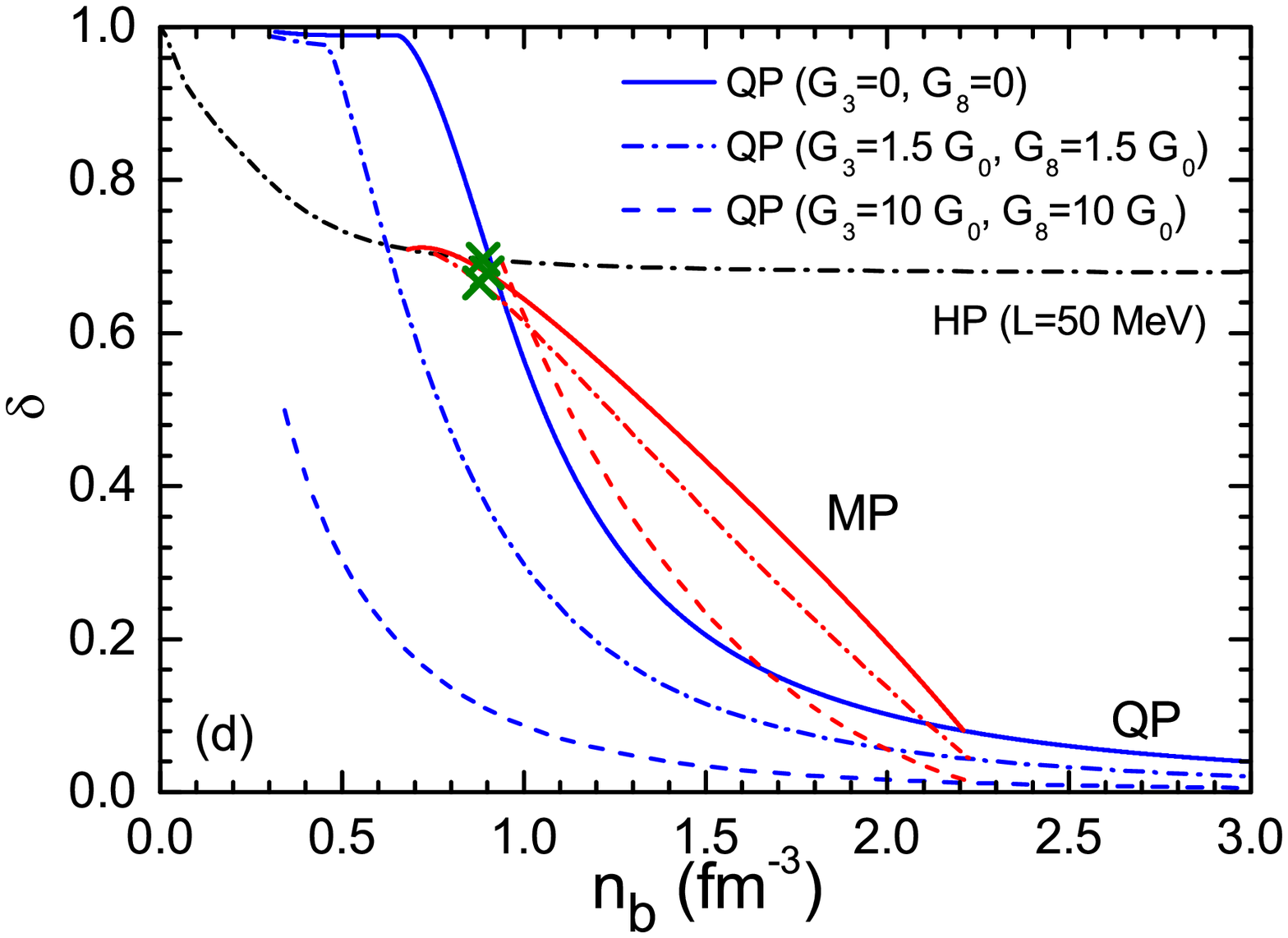} \\
\includegraphics[bb=40 5 580 400, width=7 cm,clip]{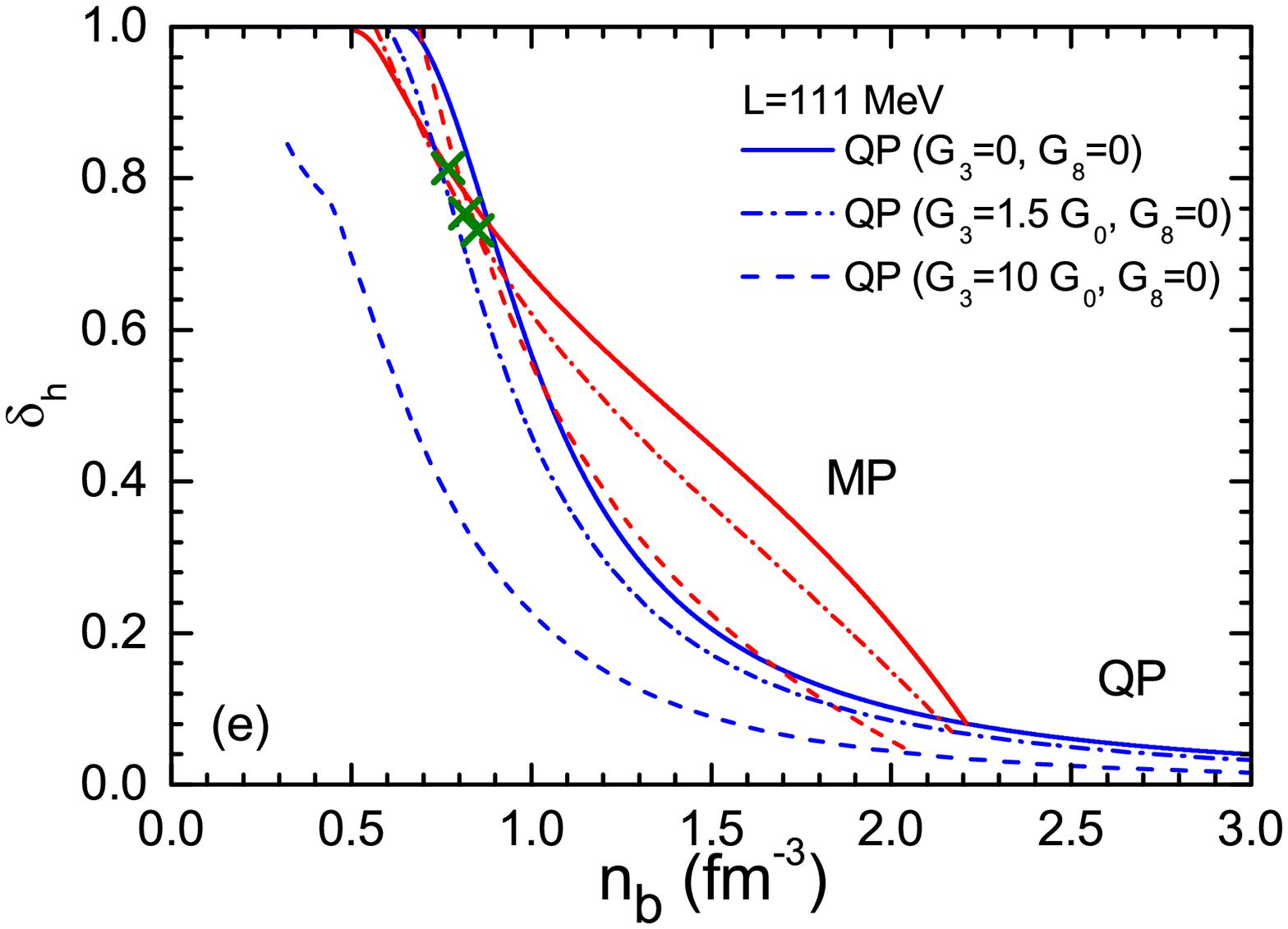}%
\includegraphics[bb=40 5 580 400, width=7 cm,clip]{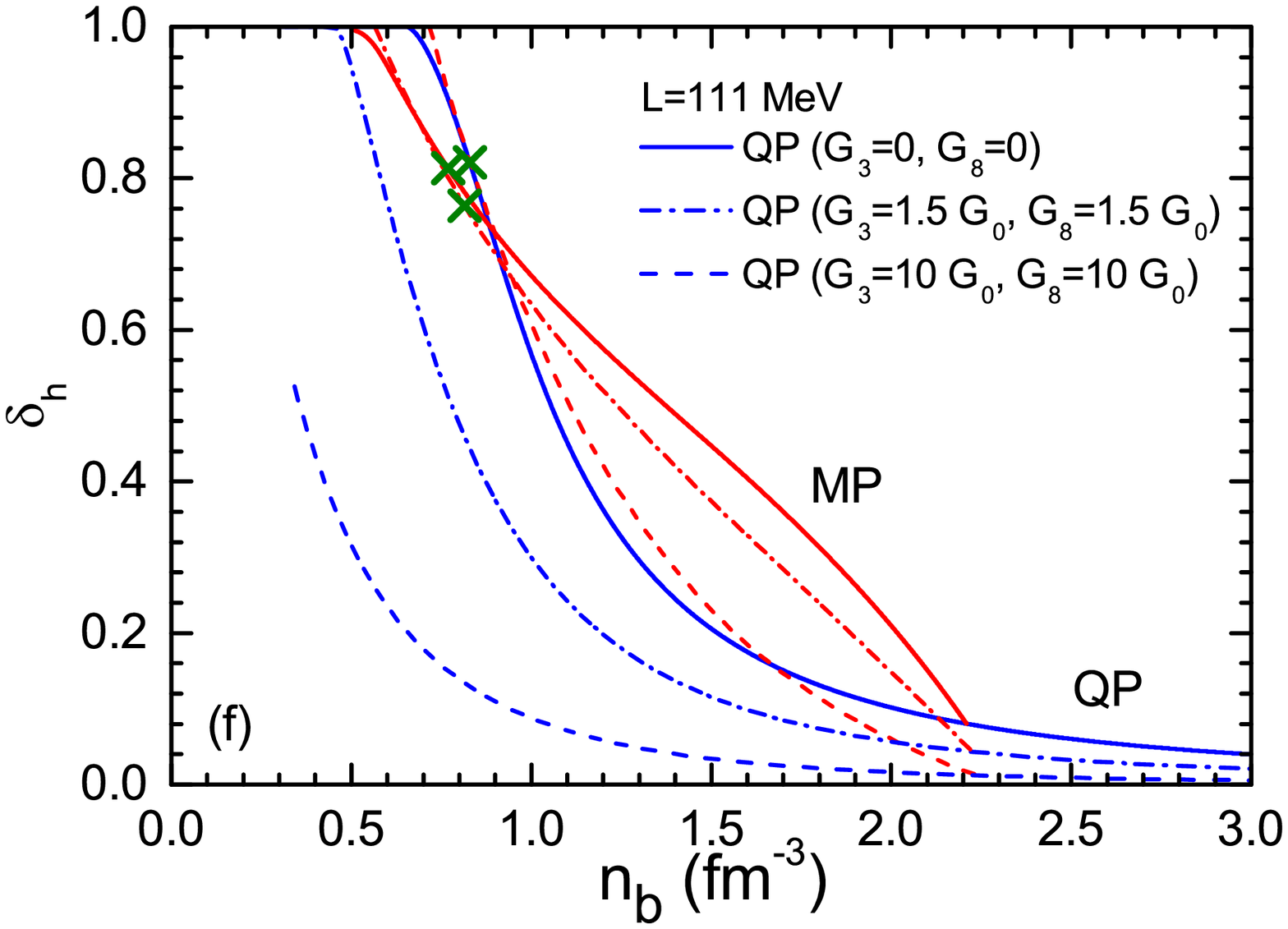} \\
\includegraphics[bb=40 5 580 400, width=7 cm,clip]{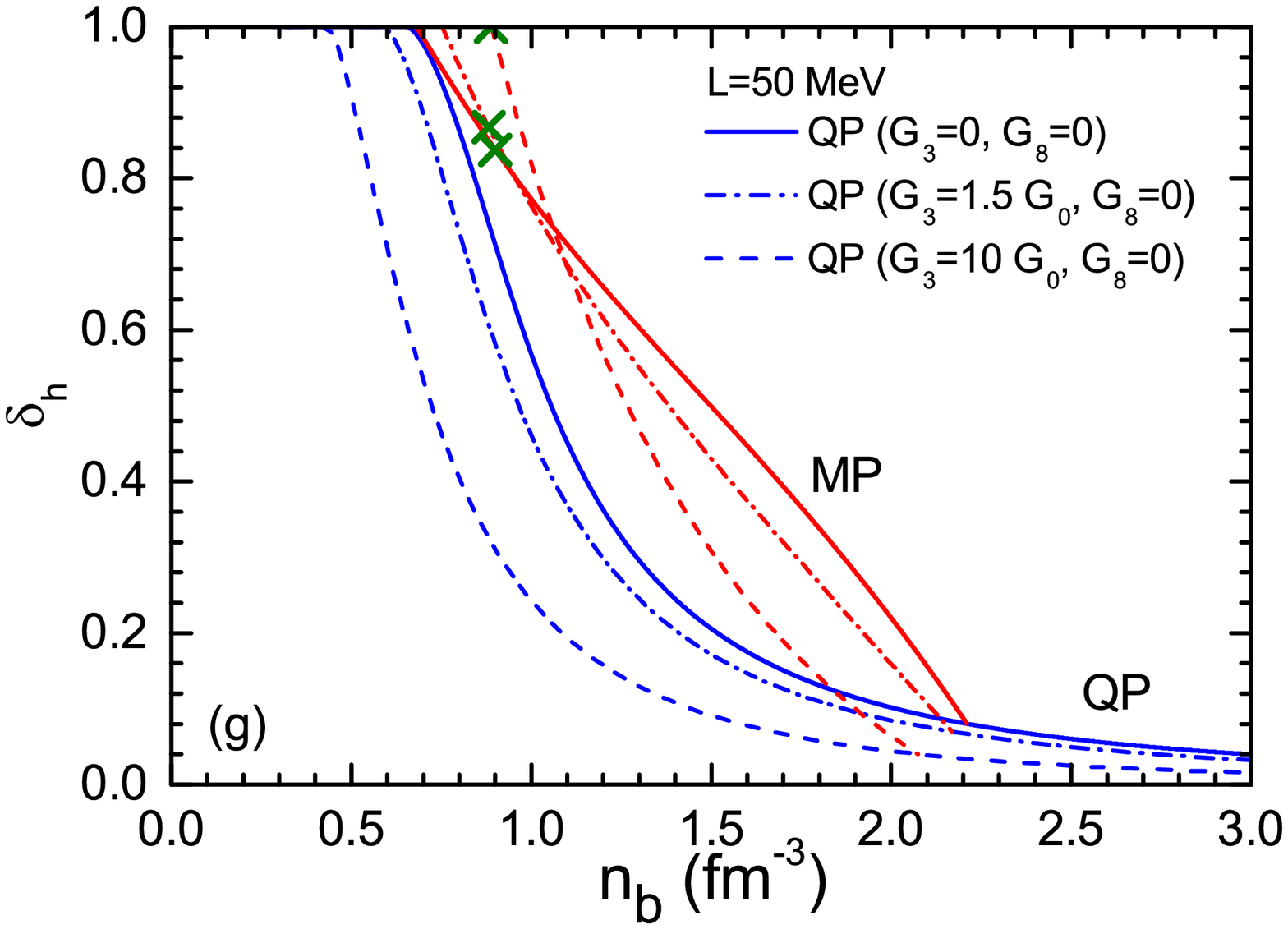}%
\includegraphics[bb=40 5 580 400, width=7 cm,clip]{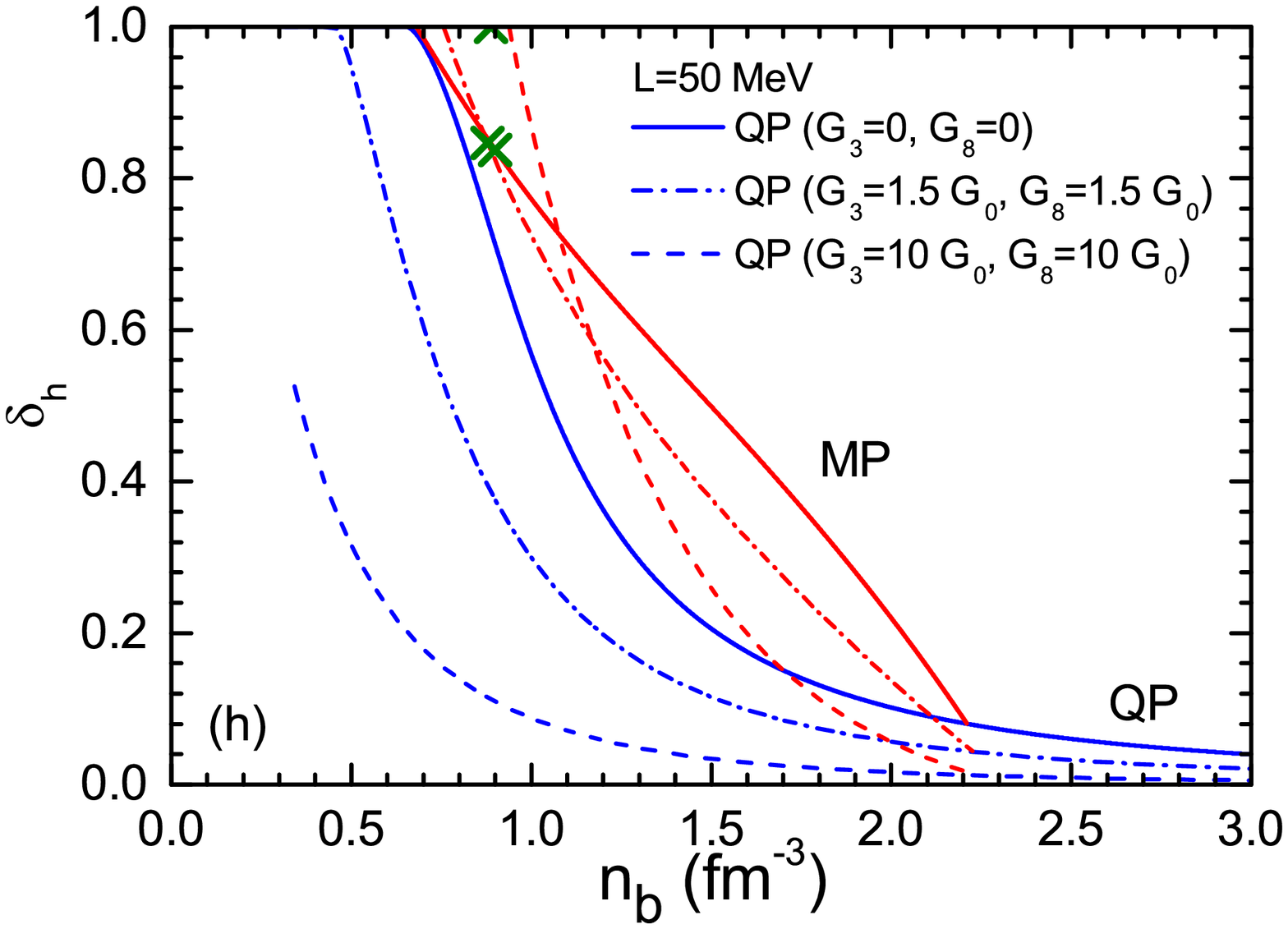} \\
\caption{(Color online) The isospin asymmetry and hypercharge fraction for 
$(G_3/G_0,G_8/G_0)=(0,0),(1.5,0),(1.5,1.5),(10,0)$, and $(10,10)$.}
\label{fig:5xdelta}
\end{figure*}

We now discuss the isovector-vector coupling ($G_3$) and hypercharge-vector ($G_8$) 
coupling effects in terms of the isospin asymmetry ($\delta$)
and hypercharge fraction ($\delta_h$).
We show $\delta$ ($\delta_h$) as a function of the baryon number density
in the top (bottom) two rows in Fig.~\ref{fig:5xdelta}.
We first find that $G_3$ and $G_8$ obviously suppress
$\delta$ and $\delta_h$ in the pure quark phase and the mixed phase.
We also note that $\delta$ decreases rapidly
when the $s$ quark appears and $\delta_h$ becomes smaller than unity.
With increasing $n_b$,
the effect from different symmetry energy slope $L$ becomes smaller.
Finite $G_3$ and $G_8$ increase the energy
and delay the appearance of the mixed phase,
i.e., two couplings push up $n_b^{(1)}$.
By comparison, $n_b^{(2)}$ is determined
by both the energy increase and the asymmetry decrease in quark matter.
The increase of the energy in quark matter pushes up $n_b^{(2)}$.
The decrease of $\delta$ and thus $\mu_e$ tends
to make coexisting hadronic matter isospin symmetric and positively charged,
disfavors the mixed phase, and pushes down $n_b^{(2)}$.
When both $G_3$ and $G_8$ are switched on,
effects from the energy increase and the asymmetry decrease
seem to cancel, and $n_b^{(2)}$ is almost the same as that without $G_3$ and $G_8$.
When only $G_3$ is switched on,
effects from the asymmetry decrease are larger than
those from the energy increase,
then $n_b^{(2)}$ decreases slightly.

\begin{figure*}[tbhp]
\includegraphics[bb=20 5 580 420, width=5 cm,clip]{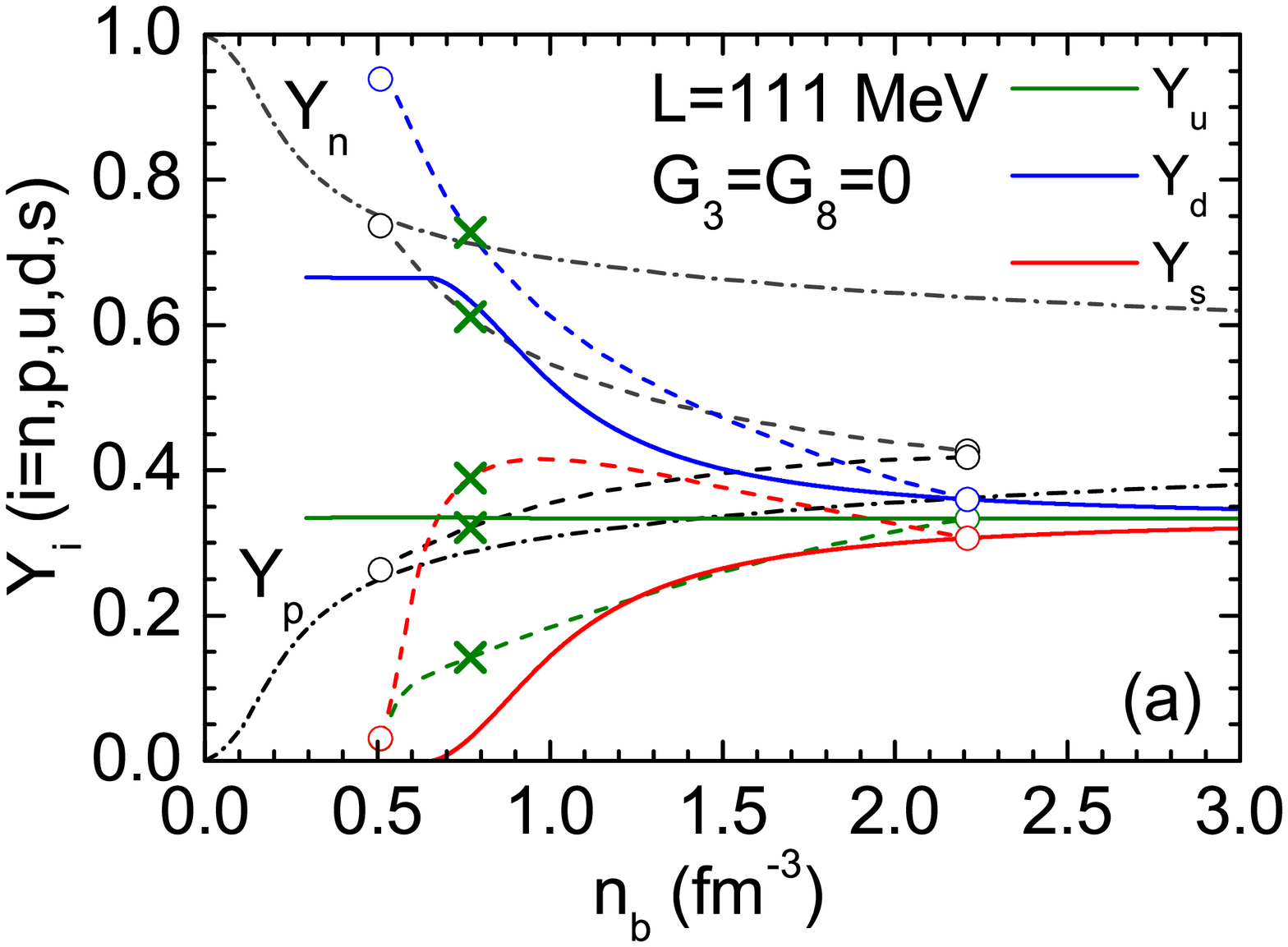}
\includegraphics[bb=20 5 580 420, width=5 cm,clip]{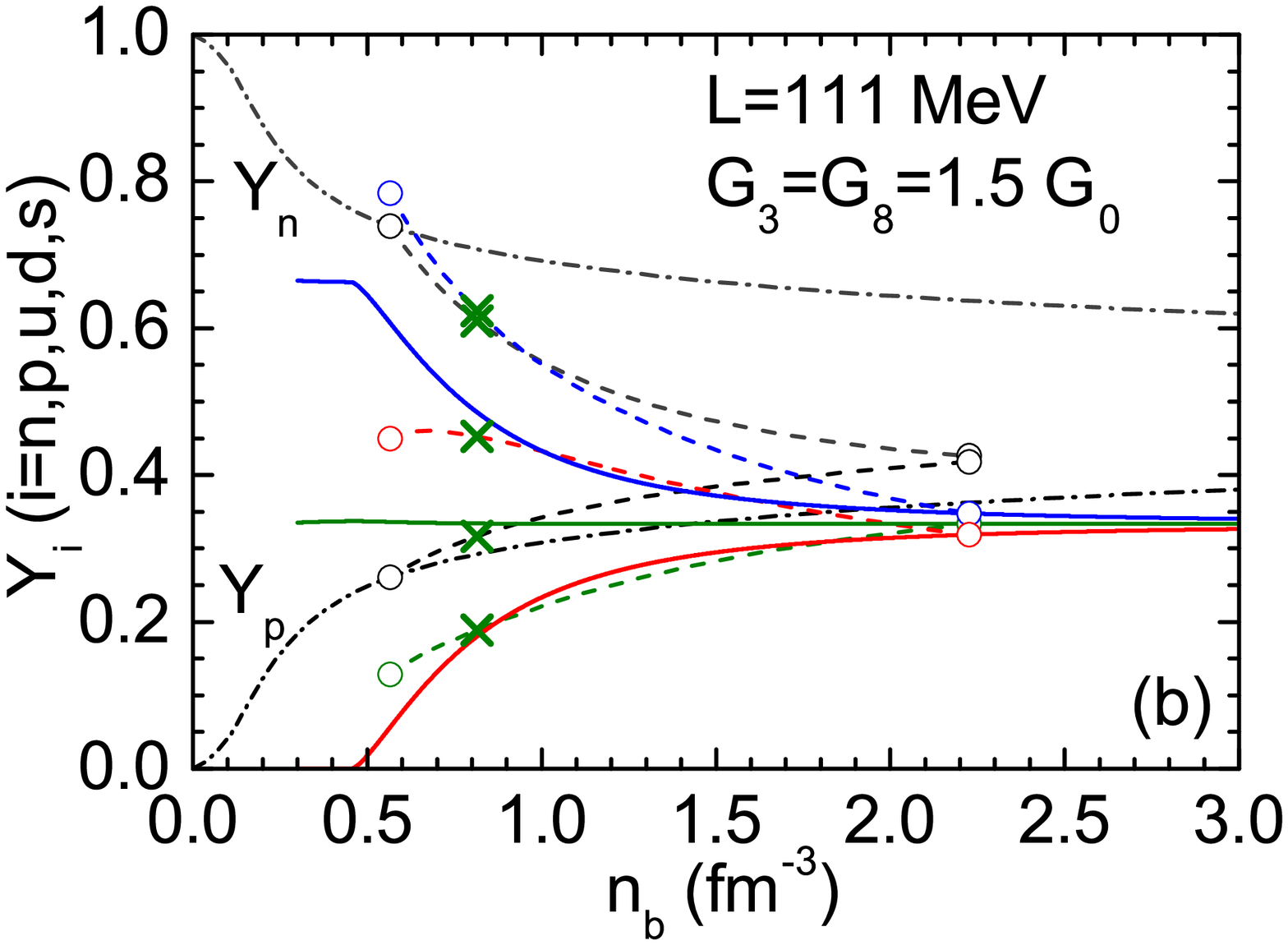}
\includegraphics[bb=20 5 580 420, width=5 cm,clip]{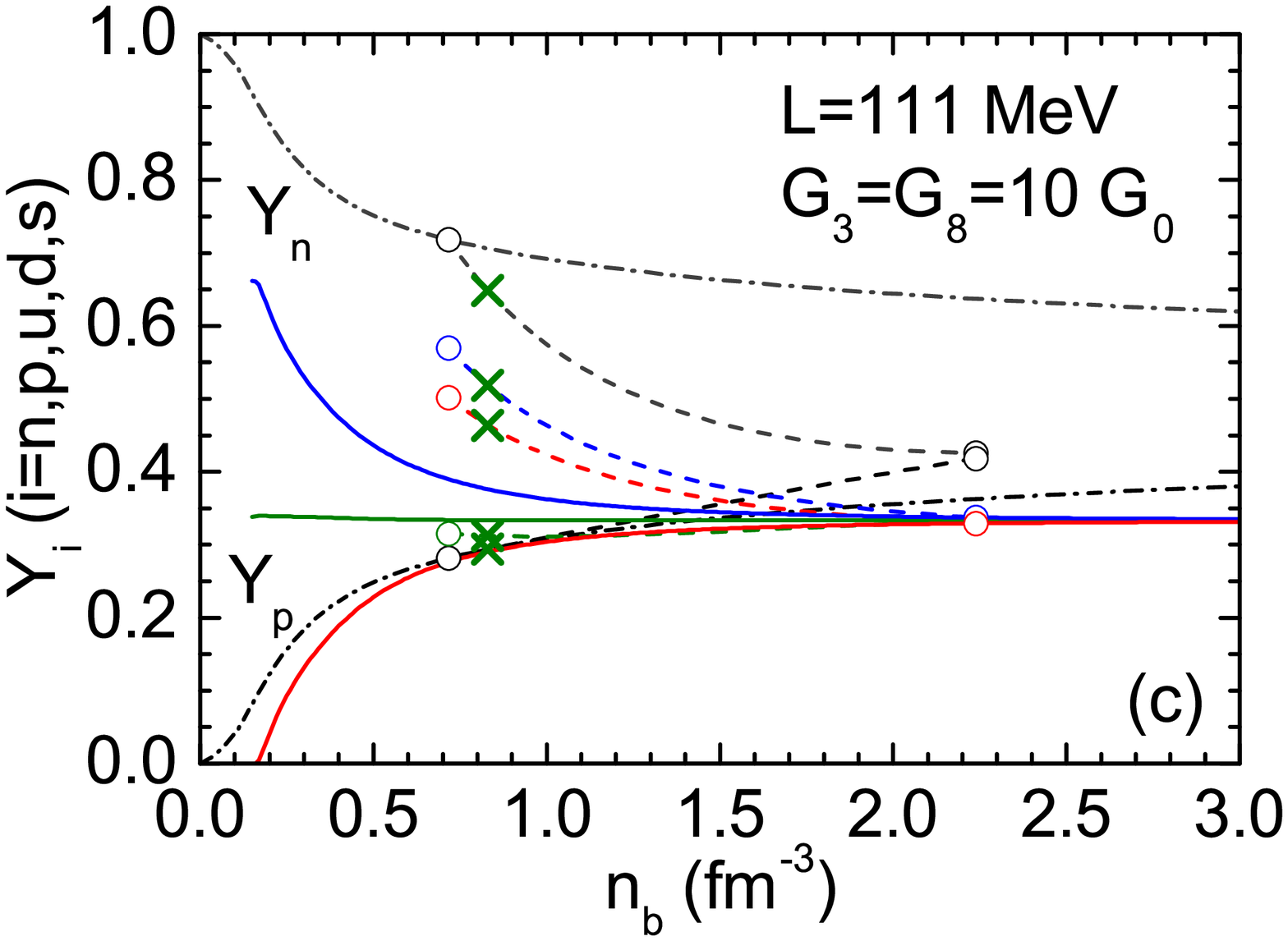}\\
\includegraphics[bb=20 5 580 420, width=5 cm,clip]{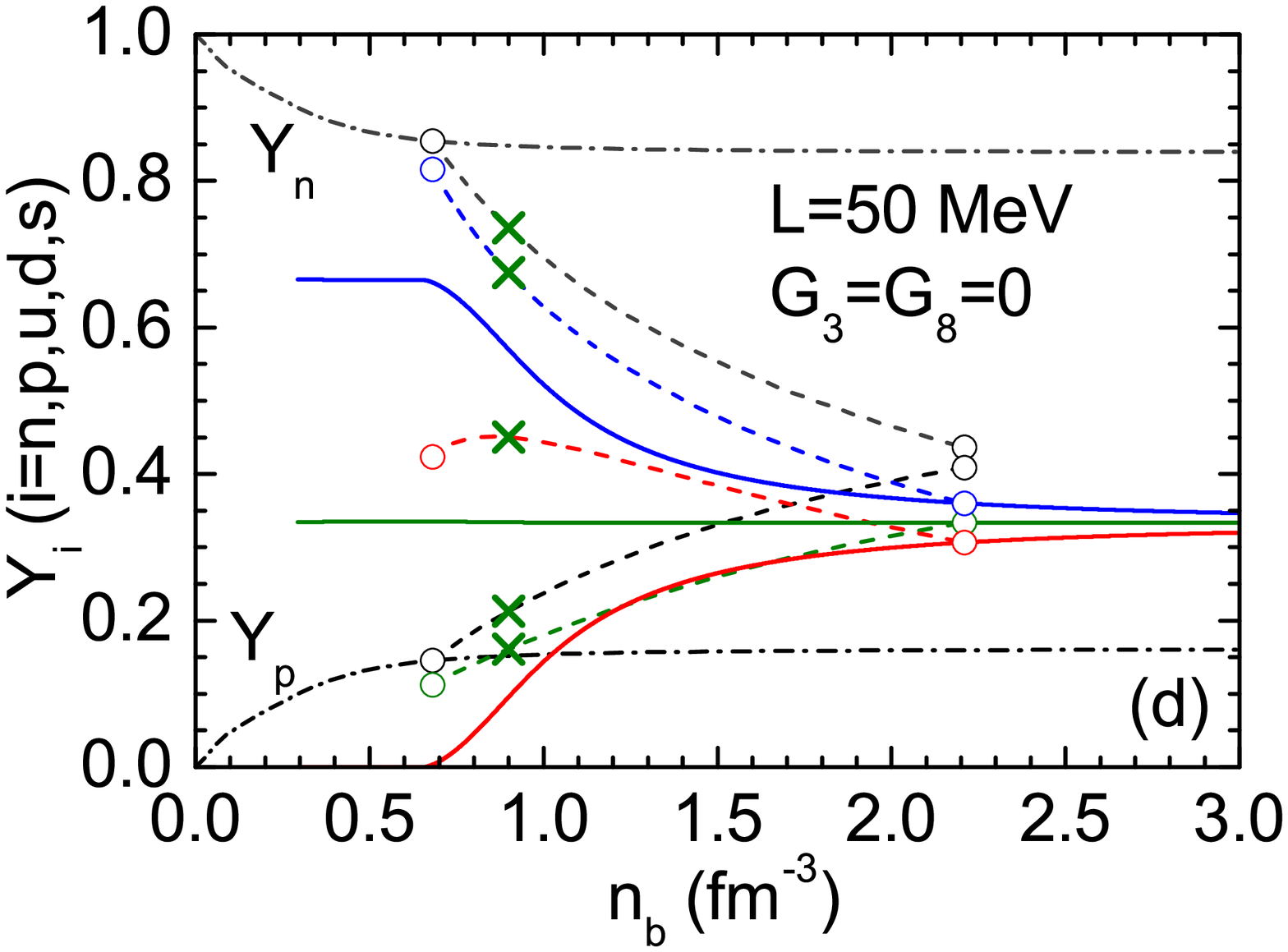}
\includegraphics[bb=20 5 580 420, width=5 cm,clip]{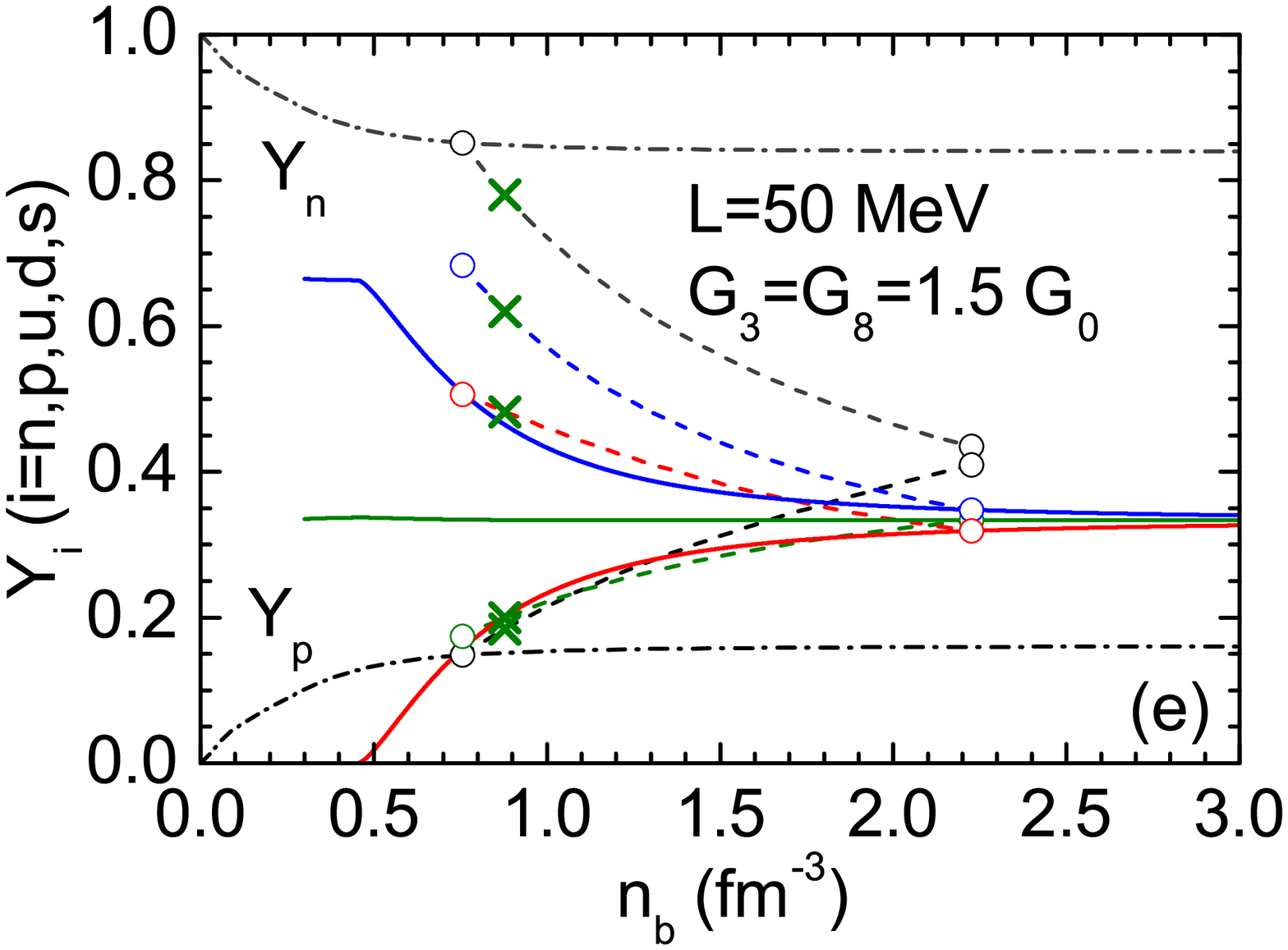}
\includegraphics[bb=20 5 580 420, width=5 cm,clip]{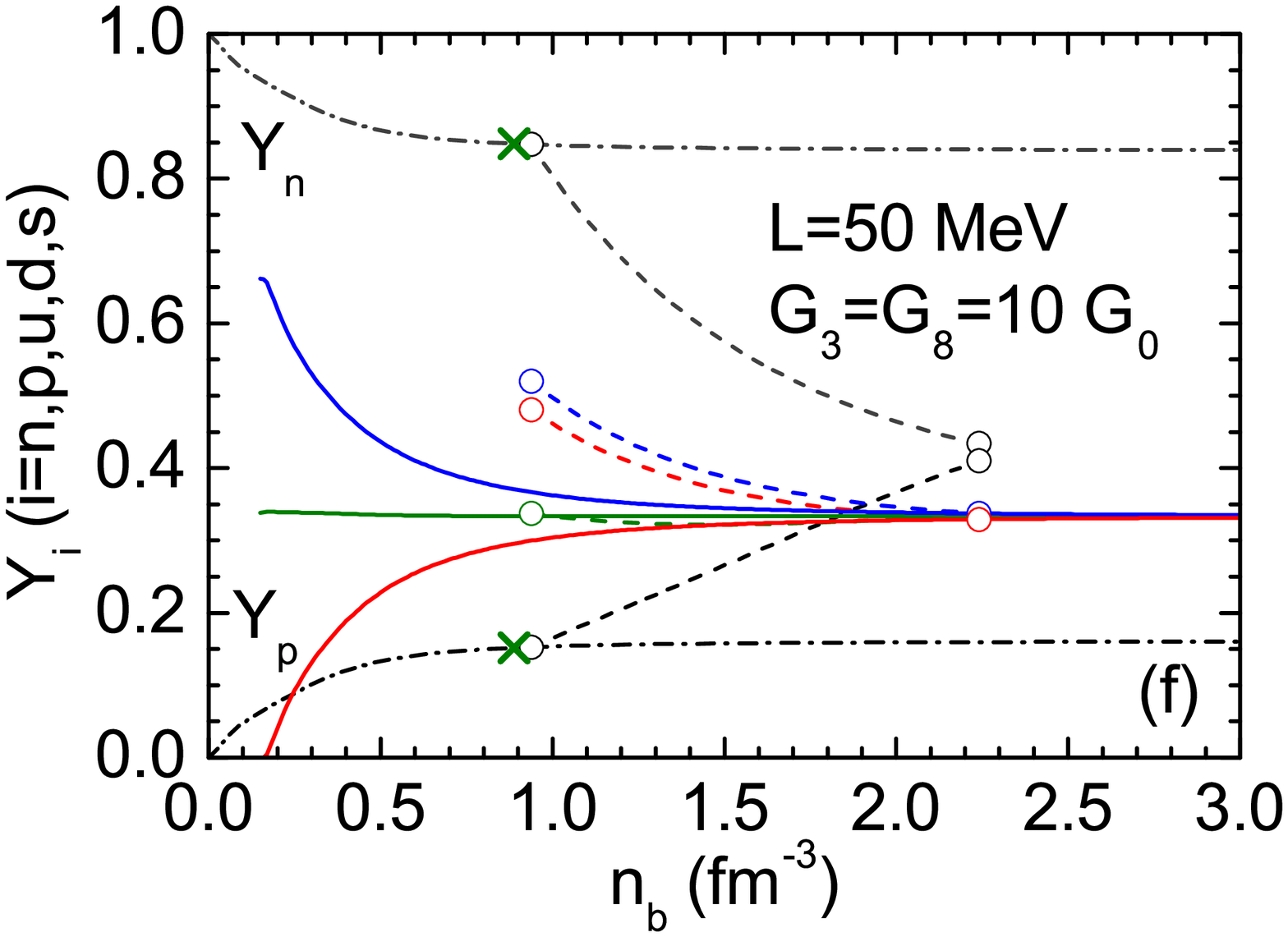}
\caption{(Color online) Particle number fractions for the pure
hadronic phase (dash-dotted lines), the mixed phase (dashed lines) and the pure quark phase (solid lines). The 
fractions of $u$, $d$, and $s$ quarks are labeled by green (middle), blue (upper), red (lower) solid lines in the 
pure quark phase and green (lower), blue (upper), red (middle) dashed lines in the mixed phase.}
\label{fig:6fraction}
\end{figure*}

Let us further discuss the effects of quark-matter symmetry energy
on particle fractions.
In Fig.~\ref{fig:6fraction}, we show the particle number fractions
in hadron matter, quark matter, and the mixed phase,
as functions of the baryon number density.
As already mentioned,
the electron chemical potential in the mixed phase is larger
than that in pure quark matter, and therefore
the fractions of $d$ and $s$ quarks in the mixed phase are
larger than those in quark matter.
When the isovector-vector coupling $G_3$ and hypercharge-vector 
coupling $G_8$ are taken into account,
the differences between quark fractions become smaller
and the quark matter tends to be more SU(3) symmetric.
When the $s$ quark appears, the differences 
between the $u$ quark and $d$ quark are suppressed furthermore.

\subsection{Neutron stars}

\begin{figure*}[htbp]
\includegraphics[bb=40 5 580 580, width=7 cm,clip]{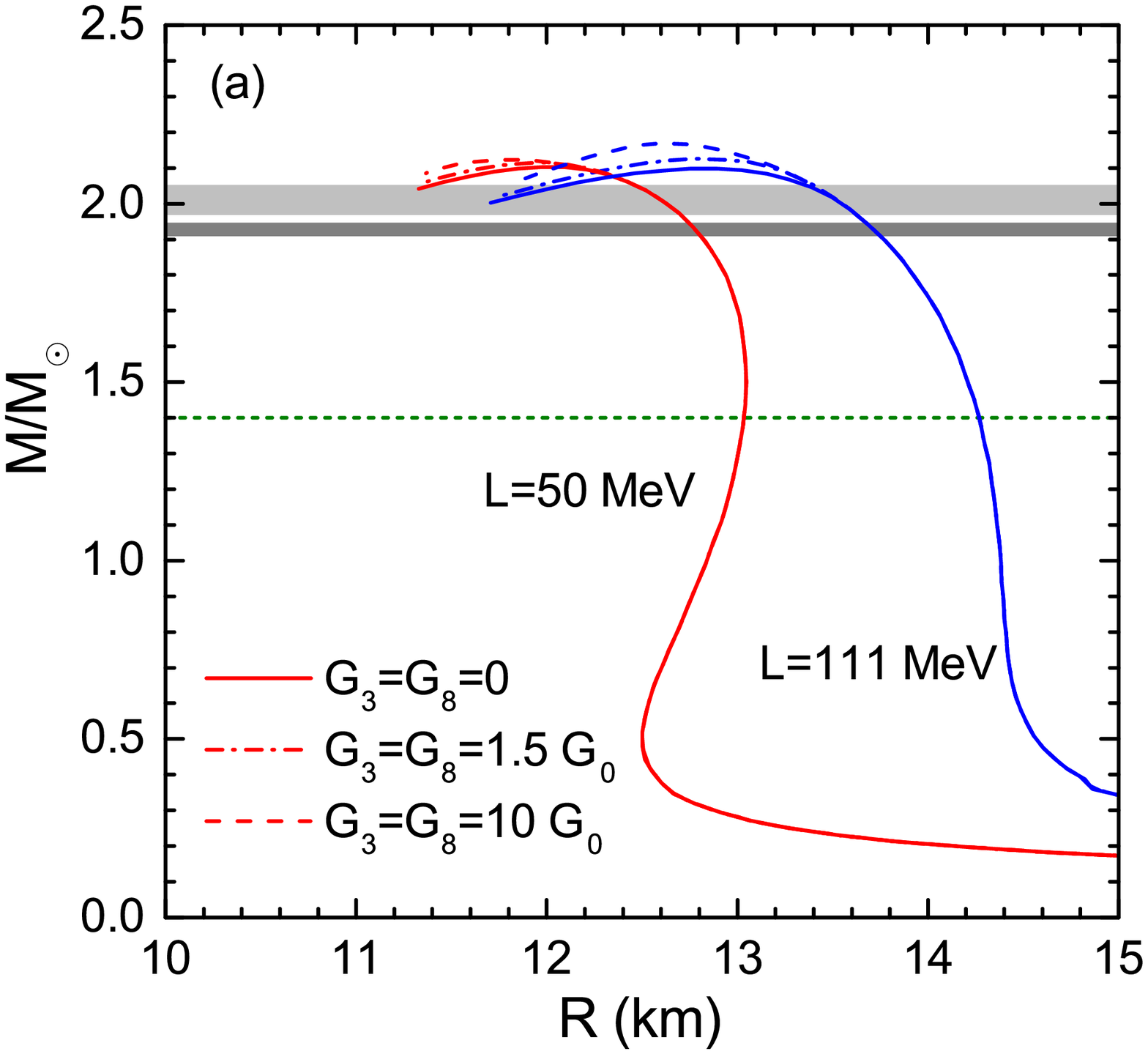}
\includegraphics[bb=40 5 580 580, width=7 cm,clip]{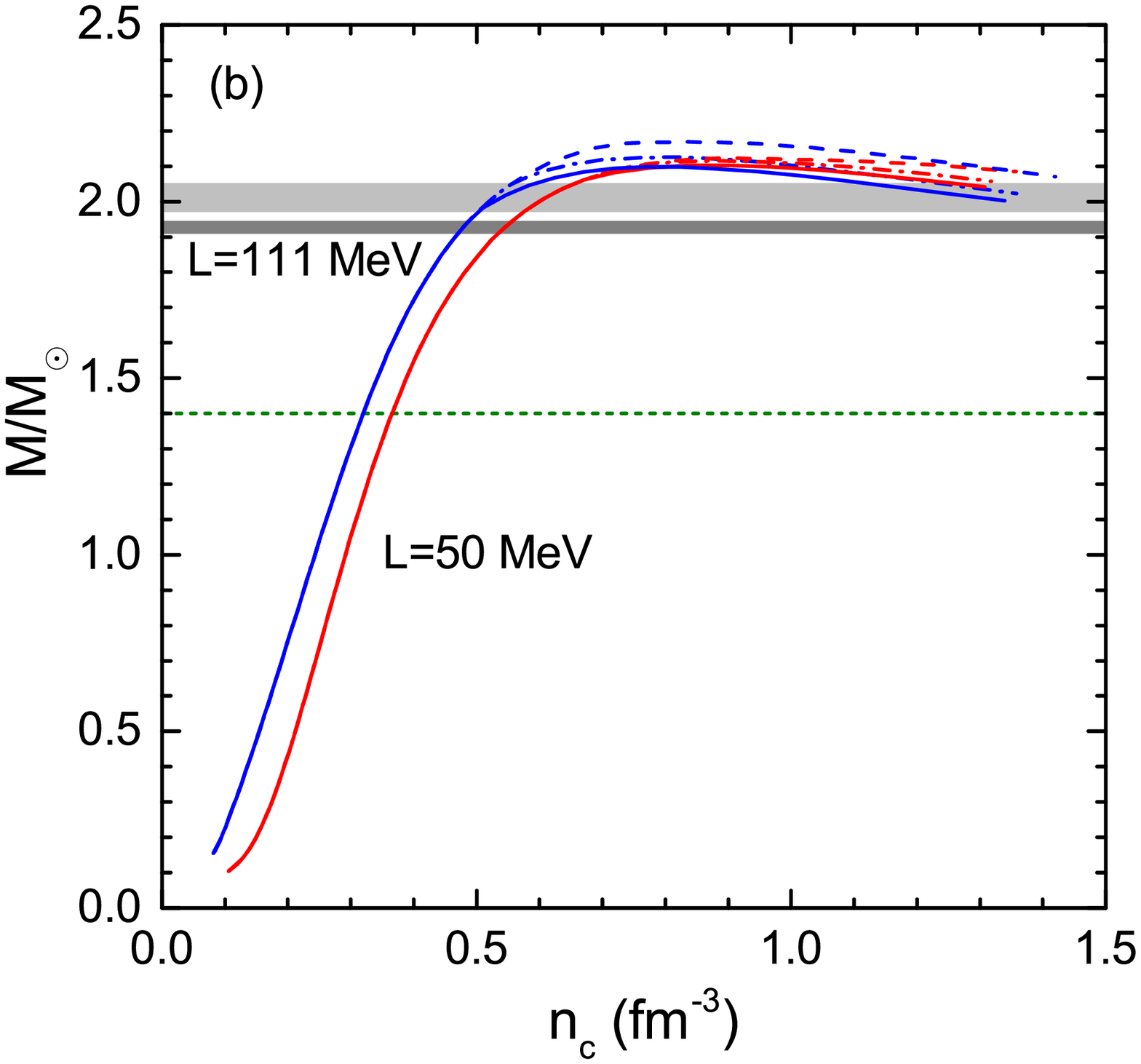}
\caption{(Color online) The left panel plots mass-radius relations
of neutron stars for different EOS. The solid lines and the dash-dotted
lines (dashed lines) show the results without and with isovector-vector coupling and hypercharge-vector coupling
respectively. The right panel shows maximum mass as a function of
neutron-star central density.}
\label{fig:7rm}
\end{figure*}

Using the EOS of pure hadronic matter, the TM1 parameter set predicts a maximum neutron-star mass
of $2.18M_{\odot}$, with $M_{\odot}$ being the solar mass.
The neutron-star observations of
PSR J1614-2230~\cite{Demo10,Fons16} and PSR J0348+0432~\cite{Anto13}
constrain that the neutron-star maximum mass needs to be larger
than $2\ M\odot$.
To examine the effect of isovector-vector coupling $G_3$ and hypercharge-vector 
coupling $G_8$ on the properties of
neutron stars, we solve the Tolman-Oppenheimer-Volkoff (TOV) equation by using
models listed in Table \ref{tab:2coex}.
The mass-radius relation is presented in the left panel of Fig.~\ref{fig:7rm}.
The right panel shows the neutron-star mass
as a function of the neutron-star central density $n_c$.
We also identified the central density of the maximum mass star in Figs.~\ref{fig:2nbp}, \ref{fig:4xmunxmue}, \ref{fig:5xdelta}, and \ref{fig:6fraction} by green cross marks, which show clearly where the central density of the maximum mass
is located. With models TM1, TM1e/NJL-V, and TM1e/NJL-VRY1, the central density may locate in the mixed phase close to the first transition density $n_b^{(1)}$, while the central density locates in the pure hadronic matter with model TM1e/NJL-VRY2.
We find that the isovector-vector coupling enhances the maximum mass 
of neutron stars slightly,
but the effect is inconspicuous.
For the results of TM1e ($L=50~\MeV$),
it shows a smaller radius than that of TM1 ($L=110.8~\MeV$).
The influence of $G_3$ on neutron-star maximum mass becomes even
smaller with TM1e. This is because
the onset of the mixed phase in TM1e is later than that in TM1.
In the right panel of Fig.~\ref{fig:7rm}, we can see that
when the transition density $n_b^{(1)}$ is close to the maximum neutron 
star central density $n_c^{M_{\rm{max}}}$ (for TM1e case),
the hadron-quark coexistence has little effect for the maximum mass neutron stars.

\section{Summary}
\label{sec:summary}
Effects of the isovector-vector and hypercharge-vector couplings
in quark matter on hadron-quark phase transition
and neutron-star properties are investigated.
In this work, we have used the RMF theory to describe hadronic matter,
and the three flavor NJL model including the
isovector-vector coupling has been used for the quark matter.
The Gibbs conditions are applied to describe the hadron-quark mixed phase.
We have found that the mixed phase shrinks with the isovector-vector coupling in quark matter,
while the mixed phase moves to higher density with both isovector-vector
and hypercharge-vector couplings included.
If only isovector-vector coupling is included in quark matter, the transition density
to the mixed phase ($n_b^{(1)}$) increases by $(0.06-0.25)~\fm^{-3}$,
and the transition density to the pure quark matter ($n_b^{(2)}$) decreases 
by $(0.04-0.14)~\fm^{-3}$. The hypercharge-vector coupling delays both $n_b^{(1)}$
and $n_b^{(2)}$.
The inclusion of the isovector-vector and hypercharge-vector couplings in quark matter has similar
effects as decreasing the symmetry energy slope $L$ in hadronic matter.
Both of them can affect the asymmetry of the system.
We have found that the isovector-vector and hypercharge-vector couplings suppress
the asymmetry of the nuclear-quark matter system.
Meanwhile, the softening of the
EOS due to the hadron-quark phase transition becomes weaker
and the neutron-star maximum mass increases without modifying
the neutron-star radius around $M \sim 1.4\ M\odot$,
where the central density is $n_b \sim(2-3)\ n_0$.

\section*{Acknowledgment}

This work was supported in part by
the Grants-in-Aid for Scientific Research from JSPS
(Grants No. 15K05079, No. 15H03663, and No. 16K05350),
the Grants-in-Aid for Scientific Research on Innovative Areas from MEXT
(Grants No. 24105001 and No. 24105008),
the National Natural Science Foundation of China (Grants No. 11675083),
and by the Yukawa International Program for Quark-hadron Sciences (YIPQS).
X. W. acknowledges the financial support provided by the China Scholarship Council (CSC).


\newpage

\end{document}